\documentclass[10pt, conference, compsocconf]{IEEEtran}
\ifCLASSINFOpdf
\else
\fi
\usepackage[cmex10]{amsmath}

\usepackage{hyperref}
\usepackage{graphicx}
\usepackage{float,amsmath,amssymb,dcolumn,bm,hyperref,array}
\usepackage[usenames,dvipsnames]{color}
\usepackage{subfig}
\usepackage[utf8]{inputenc}
\usepackage{import}

\graphicspath{{./Files/PDFs/1D/}{./Files/PDFs/2D/Hydro/}{./Files/PDFs/2D/MHD/}{./Files/PDFs/3D/Hydro/}{./Files/PDFs/3D/MHD/}{./Files/PDFs/NVIDIA/}{../Files/PDFs/1D/}{../Files/PDFs/2D/Hydro/}{../Files/PDFs/2D/MHD/}{../Files/PDFs/3D/Hydro/}{../Files/PDFs/3D/MHD/}{../Files/PDFs/NVIDIA/}}

\begin{document}
\title{Three Dimensional Pseudo-Spectral Compressible Magnetohydrodynamic GPU Code for Astrophysical Plasma Simulation}


\author{\IEEEauthorblockN{Rupak Mukherjee, R Ganesh, Vinod Saini, Udaya Maurya}
\IEEEauthorblockA{Institute for Plasma Research, HBNI,\\
Bhat, Gandhinagar - 382428, India\\
rupakmukherjee01@gmail.com; rupak@ipr.res.in\\
ganesh@ipr.res.in}\\
\and
\IEEEauthorblockN{Nagavijayalakshmi Vydyanathan, B Sharma}
\IEEEauthorblockA{NVIDIA\\
Bengaluru, 560045, India\\
nvydyanathan@nvidia.com\\
bharatk@nvidia.com}
}


%Telephone: (800) 555--1212, Fax: (888) 555--1212}


\maketitle

\begin{abstract}
This paper presents the benchmarking and scaling studies of a GPU accelerated three dimensional compressible magnetohydrodynamic code. The code is developed keeping an eye to explain the large and intermediate scale magnetic field generation is cosmos as well as in nuclear fusion reactors in the light of the theory given by Eugene Newman Parker. The spatial derivatives of the code are pseudo-spectral method based and the time solvers are explicit. GPU acceleration is achieved with minimal code changes through OpenACC parallelization and use of NVIDIA CUDA Fast Fourier Transform library (cuFFT). NVIDIA’s unified memory is leveraged to enable over-subscription of the GPU device memory for seamless out-of-core processing of large grids. Our experimental results indicate that the GPU accelerated code is able to achieve upto two orders of magnitude speedup over a corresponding OpenMP parallel, FFTW library based code, on a NVIDIA Tesla P100 GPU. For large grids that require out-of-core processing on the GPU, we see a 7x speedup over the OpenMP, FFTW based code, on the Tesla P100 GPU. We also present performance analysis of the GPU accelerated code on different GPU architectures - Kepler, Pascal and Volta.
\end{abstract}


% \fi
\IEEEpeerreviewmaketitle

\section{Introduction}
Plasmas can be viewed as a charged fluid. In stellar atmosphere, because of strong thermal radiation, the neutral atoms get ionised and form an ionised gas maintaining a quasi-neutrality in space. Thus the state of matter that dominates the structure of a young star is plasma. Because of very high kinetic energy the motion of the charged particles generate a magnetic field and thus the self-consistent motion of the plasma-fluid element gets primarily governed by the Maxwell's equations coupled with the equations of hydro-dynamics. The subject that studies the self-consistent evolution of such a magnetised plasma is known as magneto-hydro-dynamics (MHD). In laboratories around the globe, there are several Tokamak experiments being carried out to mimic the fusion process in turbulent plasmas taking place in stars. The fundamental motive of such experiments are to harness the binding energy of atoms and transform it to the form of daily usable electricity for the service of mankind. Thus the three dimensional magnetised plasma turbulence is crucial to understand the fundamentals of astro-plasmas present in sun or other young stars as well as the careful operation of complicated Tokamak reactors. 

The dynamics of a fully developed turbulent plasma present in stellar objects are fundamentally different from that of a neutral fluid and hence needs a separate careful treatment. The inherent features of this plasma turbulence are needed to be understood and properly analysed to predict any unnatural behavior taking place in our nearest star ``The Sun'' or in the Tokamak reactors run over the several laboratories. 

The large and intermediate scale magnetic field generation in sun as well as in accretion disks around any compact object (for example a neutron star or a black hole) are called ``Dynamo Effect" first identified by Eugene Newman Parker. These large or intermediate scale magnetic field primarily governs the dynamics of the charged fluid (plasma) through a time dependent Lorentz force term added in the Navier-Stokes equation, thereby adding an extra dimension (and allowing the exchange of energy between kinetic and magnetic modes), to the already known and well-explored subject ``fluid dynamics''. 

Addressing some of these issues of magnetic field generation and plasma dynamics require, efficient, well parallelized numerical solver capable of handling compressible MHD physics as well as numerical complexities. In order to simulate such a physical process, the finite difference DNS schemes falls short to cope up with the requirement of the high speed performance of the code. Pseudo-spectral technique is observed to be much more accurate and faster method than finite difference methods. We have developed a Three dimensional compressible MHD solver G-MHD3D, which considers continuity, momentum and energy equations for fluid and magnetic variables with a thermodynamic closure for pressure. Code uses conservative form using pseudo-spectral method in cartesian coordinates with GPU parallelization. The numerical solver we have developed captures similar Dynamo Effect for chaotic plasma flows. The code is expected to contribute even more for better understanding of such turbulent phenomena in cosmos.

In this paper, extensive physics benchmarking studies as well as numerical scaling studies is presented along with GPU parallelization, scalability and numerical as well as physics details. It is envisaged that this new code with its extensive physics diagnostics and GPU scalability should be able to address some of the fundamental physics issues in areas such as reconnection (a phenomena of transforming back the magnetic energy to kinetic energy of the plasma), dynamos as well as MHD turbulence.

The OpenMP parallel, FFTW based pseudo-spectral DNS code, MHD3D, has been developed in-house at IPR. The Graphics Processing Unit (GPU) version of the code (G-MHD3D) has been developed with NVIDIA and the performence results have been tested at NVIDIA clusters.

\section{Code G-MHD3D}
G-MHD3D is an OpenACC~\cite{OpenACC} parallel three dimensional compressible, viscous, resistive magnetohydrodynamic GPU code using pseudo-spectral technique to address sponteneous magnetic field generation related problems. The pesudo-spectral technique uses the cuFFT libraries~\cite{cuFFT} which are one of the fastest Fourier Transform GPU libraries available till today. This technique is applied to calculate the spatial derivatives and to evaluate the non-linear terms involved in the basic underlying equations. The time derivative is solved using multiple time solvers viz. Adams-Bashforth, Runge-Kutta 4 and Predictor-Correcter algorithms and all the solvers has been checked to provide identical results. The velocity and the magnetic field strength has been normalised with the sound speed and the Alfven speed in the system. The salient features of MHD3D and G-MHD3D are listed in Table \ref{tbl:GPU_CPU_comparison}. 

We develop a new code C-FD-2D, capable of handling compressible fluids, that uses pseudo-spectral technique for spatial discretisation with a standard de-aliasing by zero padding using the $3/2$ rule to simulate the above set of equations. The equations are evaluated in two dimensions with periodic boundary conditions in cartesian co-ordinates. We use the multi-dimensional FFTW libraries to evaluate the fourier transforms, whenever needed. The time evolution of the code is performed with Adams-Bashforth, Predictor-Corrector and Runge-Kutta 4 techniques individually and all the three techniques are found to agree in the primary benchmark results. The results in this paper are calculated using Adams-Bashforth method for temporal evolution calculation. 

Following we provide the specifications of the basic details and the salient features of our 3D MHD code both for CPU and GPU versions.

\begin{table}[h]
\begin{tabular}{ |c|c| } 
 \hline
 &\\
 Type of equations & Compressible Navier-Stokes + Maxwell\textquoteright s Equations\\
 &\\ 
 Dimensionality & 3D  \\ 
 &\\
 Spatial Derivative Solver & Pseudo-Spectral  \\
 &\\
 Time Discretization & Adams-Bashforth, Runge-Kutta 4, Predictor-Corrector\\
 &\\
 Computer architecture & CPU, GPU (NVIDIA) \\
 &\\
 Parallelization & OpenMP, OpenACC \\
 &\\
 External Libraries & FFTW, cuFFT \\
 &\\
 Precision & Double \\
 &\\
 Language & Fortran 95 \\
 &\\
 \hline
\end{tabular}
\caption{Features of MHD3D and G-MHD3D.}
\label{tbl:GPU_CPU_comparison}
\end{table}

\section{Governing equations}
The basic equations that are evolved in the code with different specific initial conditions are as follows:\\
\begin{eqnarray}
\label{Mass density} \text{Mass density:}~ && \frac{\partial \rho}{\partial t} + \vec{\nabla} \cdot \left(\rho \vec{u}\right) = 0\\
\label{Momentum equation} \text{Momentum equation:}~ && \frac{\partial (\rho \vec{u})}{\partial t} + \vec{\nabla} \cdot \left[ \rho \vec{u} \otimes \vec{u} \right] \nonumber \\
&& = \frac{\vec{J} \times \vec{B}}{c} - \vec{\nabla} P \nonumber \\
&& + \vec{\nabla} \cdot (2 \nu \rho \bar{\bar S}) + \rho \vec{F}\\
\label{Shear viscosity} \text{Shear viscosity:}~ && S_{ij} = \frac{1}{2}(\partial_i u_j + \partial_j u_i) \nonumber \\
&& - \frac{1}{3} \delta_{ij} \theta\\
\label{Dilation} \text{Dilation:}~ && \theta = \vec{\nabla} \cdot \vec{u}\\
\label{Equation of state} \text{Equation of state:}~ && P = \gamma \rho K T = C_s^2 \rho\\
\label{Non relativistic Ampere's law} \text{ Ampere's law:}~ && \vec{J} = \frac{c}{4 \pi} \vec{\nabla} \times \vec{B}\\
\label{Faraday's law} \text{Faraday's law:}~ && \frac{\partial B}{\partial t} = - c \vec{\nabla} \times \vec{E}\\
\label{Ohm's law} \text{Ohm's law:}~ && \vec{E} + \frac{\vec{u} \times \vec{B}}{c} = \frac{1}{\sigma} \vec{J}\\
\label{No magnetic monopole} \text{No magnetic monopole:}~ && \vec{\nabla} \cdot \vec{B} = 0
\end{eqnarray}
Putting \ref{Non relativistic Ampere's law} into \ref{Momentum equation} we get,
\begin{eqnarray*}
&& \frac{\partial (\rho \vec{u})}{\partial t} + \vec{\nabla} \cdot \left[ \rho \vec{u} \otimes \vec{u} + \left(P+\frac{1}{8\pi}|\vec{B}|^2\right) {\bf{I}} - \frac{1}{4\pi} \vec{B} \otimes \vec{B} \right] \\
&& = \vec{\nabla} \cdot (2 \nu \rho \bar{\bar S}) + \rho \vec{F}
\end{eqnarray*}
Putting \ref{Ohm's law} into \ref{Faraday's law} and using \ref{Non relativistic Ampere's law} \& \ref{No magnetic monopole} we get,
\begin{eqnarray*}
&& \frac{\partial \vec{B}}{\partial t} = - c \vec{\nabla} \times \left[ -\frac{\vec{u} \times \vec{B}}{c} + \frac{1}{\sigma} \vec{J}\right] \\
&& = \vec{\nabla} \times \vec{u} \times \vec{B} - \frac{c}{4 \pi \sigma} \vec{\nabla} \times \vec{\nabla} \times \vec{B} = \vec{\nabla} \times \vec{u} \times \vec{B} + \frac{c}{4 \pi \sigma} \nabla^2 \vec{B}\\
\Rightarrow && \frac{\partial \vec{B}}{\partial t} + \vec{\nabla} \cdot \left( \vec{u} \otimes \vec{B} - \vec{B} \otimes \vec{u}\right) = \frac{c}{4 \pi \sigma} \nabla^2 \vec{B} \\
\Rightarrow && \frac{\partial \vec{B}}{\partial t} + \vec{\nabla} \cdot \left( \vec{u} \otimes \vec{B} - \vec{B} \otimes \vec{u}\right) = \eta \nabla^2 \vec{B}, ~~\text{where,}~~ \eta = \frac{c}{4 \pi \sigma}
\end{eqnarray*}
The internal energy is evaluated by the time evolution of the following equation:
\begin{eqnarray*}
&& \frac{\partial E}{\partial t} + \vec{\nabla} \cdot [ \left( E + P_{tot} \right)\vec{u} - \frac{1}{4\pi} \vec{u}\cdot\left( \vec{B} \otimes \vec{B} \right) \\
&& - 2 \nu \rho \vec{u} \cdot \bar{\bar S} - \frac{\eta}{4\pi} \vec{B} \times \left(\vec{\nabla} \times \vec{B} \right) ] = 0
\end{eqnarray*}
where $E = \rho \epsilon_{int} + \frac{1}{2}\rho|\vec{u}|^2 + \frac{1}{8\pi}|\vec{B}|^2$ and $P_{tot} = P + \frac{1}{8\pi}|\vec{B}|^2$.\\

Thus the complete set of MHD equtations are \\
\begin{eqnarray}
&& \label{MHD1} \frac{\partial \rho}{\partial t} + \vec{\nabla} \cdot \left(\rho \vec{u}\right) = 0\\
&& \label{MHD2} \frac{\partial (\rho \vec{u})}{\partial t} + \vec{\nabla} \cdot \left[ \rho \vec{u} \otimes \vec{u} + \left(P+\frac{1}{8\pi}|\vec{B}|^2\right) {\bf{I}} - \frac{1}{4\pi} \vec{B} \otimes \vec{B} \right] \nonumber \\
&& = \vec{\nabla} \cdot (2 \nu \rho \bar{\bar S}) + \rho \vec{F}\\
&& \label{MHD3} \frac{\partial E}{\partial t} + \vec{\nabla} \cdot [ \left( E + P_{tot} \right)\vec{u} - \frac{1}{4\pi} \vec{u}\cdot\left( \vec{B} \otimes \vec{B} \right) \nonumber\\
&& - 2 \nu \rho \vec{u} \cdot \bar{\bar S} - \frac{\eta}{4\pi} \vec{B} \times \left(\vec{\nabla} \times \vec{B} \right) ] = 0\\
&& \label{MHD4} \frac{\partial \vec{B}}{\partial t} + \vec{\nabla} \cdot \left( \vec{u} \otimes \vec{B} - \vec{B} \otimes \vec{u}\right) = \eta \nabla^2 \vec{B}
\end{eqnarray}

Hence the complete set of equations in component form for a three dimensional fluid becomes,
\begin{eqnarray*}
&& \frac{\partial \rho}{\partial t} + \frac{\partial}{\partial x} (\rho u_x) + \frac{\partial}{\partial y} (\rho u_y) + \frac{\partial}{\partial z} (\rho u_z) = 0
\end{eqnarray*}
\begin{eqnarray*}
&& \frac{\partial E}{\partial t} + \frac{\partial}{\partial x} [ \left( E + P_{tot} \right)u_x \\
&& - \frac{1}{4\pi} \left( u_x B_x B_x + u_y B_x B_y + u_z B_x B_z \right) \\
&& - 2 \nu \rho \left( u_x S_{xx} + u_y S_{xy} + u_z S_{xz} \right) \\
&& - \frac{\eta}{4\pi} \left\{ B_y \left( \frac{\partial B_y}{\partial x} - \frac{\partial B_x}{\partial y} \right) + B_z \left( \frac{\partial B_z}{\partial x} - \frac{\partial B_x}{\partial B_z} \right) \right\} ] \\
&& + \frac{\partial}{\partial y} [ \left( E + P_{tot} \right)u_y \\
&& - \frac{1}{4\pi} \left( u_x B_y B_x + u_y B_y B_y + u_z B_y B_z \right) \\
&& - 2 \nu \rho \left( u_x S_{yx} + u_y S_{yy} + u_z S_{yz} \right) \\
&& + \frac{\eta}{4\pi} \left\{ B_x \left( \frac{\partial B_y}{\partial x} - \frac{\partial B_x}{\partial y} \right) - B_z \left( \frac{\partial B_z}{\partial y} - \frac{\partial B_y}{\partial B_z} \right) \right\} ] \\
&& + \frac{\partial}{\partial z} [ \left( E + P_{tot} \right)u_z \\
&& - \frac{1}{4\pi} \left( u_x B_z B_x + u_y B_z B_y + u_z B_z B_z \right) \\
&& - 2 \nu \rho \left( u_x S_{zx} + u_y S_{zy} + u_z S_{zz} \right)  \\
&& + \frac{\eta}{4\pi} \left\{ B_x \left( \frac{\partial B_z}{\partial x} - \frac{\partial B_x}{\partial z} \right) + B_y \left( \frac{\partial B_z}{\partial y} - \frac{\partial B_y}{\partial B_z} \right) \right\} ] = 0
\end{eqnarray*}
\begin{eqnarray*}
&& \frac{\partial \left( \rho u_x \right)}{\partial t} + \frac{\partial}{\partial x} \left[ \rho u_x u_x + P_{tot} - \frac{1}{4\pi} B_x B_x - 2 \nu \rho S_{xx}\right] \\
&& ~~~~~~~~~~ + \frac{\partial}{\partial y} \left[ \rho u_x u_y - \frac{1}{4\pi} B_x B_y - 2 \nu \rho S_{xy} \right] \\
&& ~~~~~~~~~~ + \frac{\partial}{\partial z} \left[ \rho u_x u_z - \frac{1}{4\pi} B_x B_z - 2 \nu \rho S_{xz}\right] = \rho F_x \\
\end{eqnarray*}
\begin{eqnarray*}
&& \frac{\partial \left( \rho u_y \right)}{\partial t} + \frac{\partial}{\partial x} \left[ \rho u_y u_x - \frac{1}{4\pi} B_y B_x - 2 \nu \rho S_{yx} \right] \\
&& ~~~~~~~~~~ + \frac{\partial}{\partial y} \left[ \rho u_y u_y + P_{tot}  - \frac{1}{4\pi} B_y B_y - 2 \nu \rho S_{yy} \right] \\
&& ~~~~~~~~~~ + \frac{\partial}{\partial z} \left[ \rho u_y u_z - \frac{1}{4\pi} B_y B_z - 2 \nu \rho S_{yz} \right] = \rho F_y\\
\end{eqnarray*}
\begin{eqnarray*}
&& \frac{\partial \left( \rho u_z \right)}{\partial t} + \frac{\partial}{\partial x} \left[ \rho u_z u_x - \frac{1}{4\pi} B_z B_x - 2 \nu \rho S_{zx} \right] \\
&& ~~~~~~~~~~ + \frac{\partial}{\partial y} \left[ \rho u_z u_y - \frac{1}{4\pi} B_z B_y - 2 \nu \rho S_{zy} \right] \\
&& ~~~~~~~~~~ + \frac{\partial}{\partial z} \left[ \rho u_z u_z + P_{tot} - \frac{1}{4\pi} B_z B_z - 2 \nu \rho S_{zz}\right] = \rho F_z
\end{eqnarray*}

\begin{eqnarray*}
&& \frac{\partial B_x}{\partial t} - \left[ \frac{\partial}{\partial y} \left( u_x B_y - u_y B_x \right) + \frac{\partial}{\partial z} \left( u_x B_z - u_z B_x \right) \right] \\
&& = \eta \left( \frac{\partial^2 B_x}{\partial x^2} + \frac{\partial^2 B_x}{\partial y^2} + \frac{\partial^2 B_x}{\partial z^2} \right)\\
&& \frac{\partial B_y}{\partial t} + \left[ \frac{\partial}{\partial x} \left( u_x B_y - u_y B_x \right) - \frac{\partial}{\partial z} \left( u_y B_z - u_z B_y \right) \right] \\
&& = \eta \left( \frac{\partial^2 B_y}{\partial x^2} + \frac{\partial^2 B_y}{\partial y^2} + \frac{\partial^2 B_y}{\partial z^2} \right)\\
&& \frac{\partial B_z}{\partial t} + \left[ \frac{\partial}{\partial x} \left( u_x B_z - u_z B_x \right) + \frac{\partial}{\partial y} \left( u_y B_z - u_z B_y \right) \right] \\
&& = \eta \left( \frac{\partial^2 B_z}{\partial x^2} + \frac{\partial^2 B_z}{\partial y^2} + \frac{\partial^2 B_z}{\partial z^2} \right)
\end{eqnarray*}
\begin{eqnarray*}
&& P_{tot} = P_{th} + \frac{1}{8\pi} (B_x^2 + B_y^2 + B_z^2); ~~ P_{th} = C_s^2 \rho \\
&& S_{xx} = \frac{du_x}{dx} - \frac{1}{3} \left( \frac{du_x}{dx} + \frac{du_y}{dy} + \frac{du_z}{dz} \right);\\ 
&& S_{xy} = \frac{1}{2} \left( \frac{du_y}{dx} + \frac{du_x}{dy} \right); ~~ S_{xz} = \frac{1}{2} \left( \frac{du_z}{dx} + \frac{du_x}{dz} \right)\\
&& S_{yx} = \frac{1}{2} \left( \frac{du_x}{dy} + \frac{du_y}{dx} \right);\\
&& S_{yy} = \frac{du_y}{dy} - \frac{1}{3} \left( \frac{du_x}{dx} + \frac{du_y}{dy} + \frac{du_z}{dz} \right);\\ && S_{yz} = \frac{1}{2} \left( \frac{du_z}{dy} + \frac{du_y}{dz} \right)\\
&& S_{zx} = \frac{1}{2} \left( \frac{du_x}{dz} + \frac{du_z}{dx} \right); ~~ S_{zy} = \frac{1}{2} \left( \frac{du_y}{dz} + \frac{du_z}{dy} \right);\\
&& S_{zz} = \frac{du_z}{dz} - \frac{1}{3} \left( \frac{du_x}{dx} + \frac{du_y}{dy} + \frac{du_z}{dz} \right) 
\end{eqnarray*}
The forcing determinates a spatial averaged velocity of the system $(\textless|\vec{U}(x,y,z,t)|\textgreater_{x,y,z})$ in the presence of dissipation ($\nu \neq 0$). We determine a reference scale of pressure ($P_{ref}$) and density ($\rho_{ref}$) to determine the reference sound speed ($C_{s_{ref}}$). This provides a sonic Mach Number at steady state $\left(M_s = \frac{\textless|\vec{U}(x,y,z,t)|\textgreater_{x,y,z}}{C_{s_{ref}}}\right)$. We also evaluate a dynamic sound speed, $C_s(t) = \sqrt{\frac{\gamma \textless P(x,y,z,t) \textgreater_{x,y,z}}{\textless \rho(x,y,z,t) \textgreater_{x,y,z}}}$. The Alfven speed is determined by $V_A(t) = \frac{\textless |\vec{B}(x,y,z,t)| \textgreater_{x,y,z}}{4 \pi \sqrt{\textless \rho(x,y,z,t) \textgreater_{x,y,z}}}$. Thus we evaluate the dynamic Alfven Mach number as $M_A(t) = \frac{\textless |\vec{U}(x,y,z,t)| \textgreater_{x,y,z}}{V_A(t)}$.\\

\subsection{Normalisation of the MHD equations}
Below we provide the normalisation of MHD equations for a special case. We consider the equation of state as $P = C_s^2 \rho$. Here we deal with the equations of unforced turbulence i.e. $F = 0$. For simplicity, we also  assume, the off-diagonal components of the viscosity term to be zero. For the sake of convenience, we consider the Lorentz force term is absent in the momentum equation.
The reduced set of equations are 
\begin{eqnarray}
&& \label{RMHD1} \frac{\partial \rho}{\partial t} + \vec{\nabla} \cdot \left(\rho \vec{u}\right) = 0\\
&& \label{RMHD2} \frac{\partial \vec{u}}{\partial t} + \left( \vec{u} \cdot \vec{\nabla} \right) \vec{u} = - C_s^2 \frac{\vec{\nabla} \rho}{\rho} + \frac{\nu}{\rho} \nabla^2 \vec{u}\\
&& \label{RMHD3} \frac{\partial \vec{B}}{\partial t} + \vec{\nabla} \times \vec{u} \times \vec{B} = \eta \nabla^2 \vec{B}
\end{eqnarray}
Define $\rho = \rho_0 \tilde{\rho}$, $t = t_0 \tilde{t}$ and $\vec{r} = L \vec{\tilde{r}}$ where, $L$ is the characteristic length-scale and $t_0$ is a characteristic timescale.\\
Eq. \ref{RMHD1} can be written as 
\begin{eqnarray*}
\frac{\rho_0}{t_0}\frac{\partial \tilde{\rho}}{\partial \tilde{t}} + \frac{\rho_0 u_0}{L}\vec{\tilde{\nabla}} \cdot \left(\tilde{\rho} \vec{\tilde{u}}\right) = 0
\end{eqnarray*}
Defining, $u_0 = \frac{L}{t_0}$, we get, 
\begin{eqnarray*}
\frac{\partial \tilde{\rho}}{\partial \tilde{t}} + \vec{\tilde{\nabla}} \cdot \left(\tilde{\rho} \vec{\tilde{u}}\right) = 0
\end{eqnarray*}
Eq. \ref{RMHD2} can be written as,
\begin{eqnarray*}
&& \frac{L}{t_0^2}\frac{\partial \vec{\tilde{u}}}{\partial \tilde{t}} + \frac{L}{t_0^2}\left( \vec{\tilde{u}} \cdot \vec{\tilde{\nabla}} \right) \vec{\tilde{u}} = - \frac{C_s^2}{L} \frac{\vec{\tilde{\nabla}} \tilde{\rho}}{\tilde{\rho}} + \frac{\nu}{\rho_0 L t_0} \frac{\tilde{\nabla}^2 \vec{\tilde{u}}}{\tilde{\rho}}\\
\Rightarrow && \frac{\partial \vec{\tilde{u}}}{\partial \tilde{t}} + \left( \vec{\tilde{u}} \cdot \vec{\tilde{\nabla}} \right) \vec{\tilde{u}} = - \frac{C_s^2}{u_0^2} \frac{\vec{\tilde{\nabla}} \tilde{\rho}}{\tilde{\rho}} + \frac{\nu}{\rho_0 L u_0} \frac{\tilde{\nabla}^2 \vec{\tilde{u}}}{\tilde{\rho}}\\
\Rightarrow && \frac{\partial \vec{\tilde{u}}}{\partial \tilde{t}} + \left( \vec{\tilde{u}} \cdot \vec{\tilde{\nabla}} \right) \vec{\tilde{u}} = - \frac{1}{M_s^2} \frac{\vec{\tilde{\nabla}} \tilde{\rho}}{\tilde{\rho}} + \frac{1}{Re} \frac{\tilde{\nabla}^2 \vec{\tilde{u}}}{\tilde{\rho}}
\end{eqnarray*}
where, $M_s = \frac{u_0}{C_s}$ and $Re = \frac{\rho_0 u_0 L}{\nu}$\\
Eq. \ref{RMHD3} can be written as 
\begin{eqnarray*}
&& \frac{B_0}{t_0}\frac{\partial \vec{\tilde{B}}}{\partial \tilde{t}} + \frac{B_0}{t_0}\vec{\tilde{\nabla}} \times \vec{\tilde{u}} \times \vec{\tilde{B}} = \frac{\eta B_0}{L^2} \tilde{\nabla}^2 \vec{\tilde{B}}\\
\Rightarrow && \frac{\partial \vec{\tilde{B}}}{\partial \tilde{t}} + \vec{\tilde{\nabla}} \times \vec{\tilde{u}} \times \vec{\tilde{B}} = \frac{\eta t_0}{L^2} \tilde{\nabla}^2 \vec{\tilde{B}}\\
\Rightarrow && \frac{\partial \vec{\tilde{B}}}{\partial \tilde{t}} + \vec{\tilde{\nabla}} \times \vec{\tilde{u}} \times \vec{\tilde{B}} = \frac{\eta}{L u_0} \tilde{\nabla}^2 \vec{\tilde{B}}\\
\Rightarrow && \frac{\partial \vec{\tilde{B}}}{\partial \tilde{t}} + \vec{\tilde{\nabla}} \times \vec{\tilde{u}} \times \vec{\tilde{B}} = \frac{1}{R_m} \tilde{\nabla}^2 \vec{\tilde{B}}
\end{eqnarray*}
where, $R_m = \frac{L u_0}{\eta}$.\\
Removing the $``\tilde{~}"$ sign from the variables, we write the equations as:
\begin{eqnarray*}
&& \frac{\partial \rho}{\partial t} + \vec{\nabla} \cdot \left(\rho \vec{u}\right) = 0\\
&& \frac{\partial \vec{u}}{\partial t} + \left( \vec{u} \cdot \vec{\nabla} \right) \vec{u} = - \frac{1}{M_s^2} \frac{\vec{\nabla} \rho}{\rho} + \frac{1}{Re} \frac{\nabla^2 \vec{u}}{\rho}\\
&& \frac{\partial \vec{B}}{\partial t} + \vec{\nabla} \times \vec{u} \times \vec{B} = \frac{1}{R_m} \nabla^2 \vec{B}
\end{eqnarray*}
This normalisation is followed throughout the code.

\section{Two dimensional numerical tests}

\subsection{Incompressible hydrodynamic flow}
\begin{figure}[h!]
\begin{center}
\subfloat[]{\includegraphics[scale=0.345]{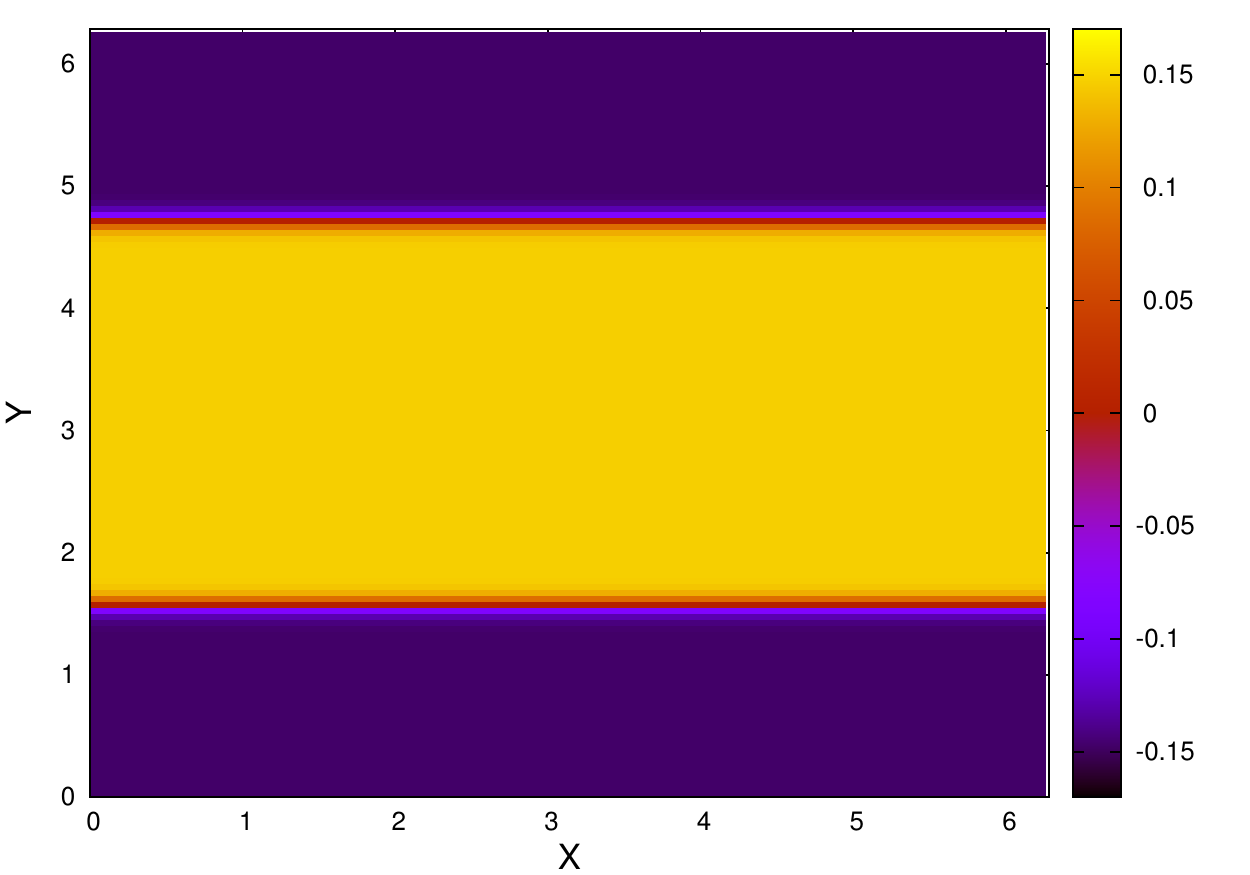}}
\subfloat[]{\includegraphics[scale=0.345]{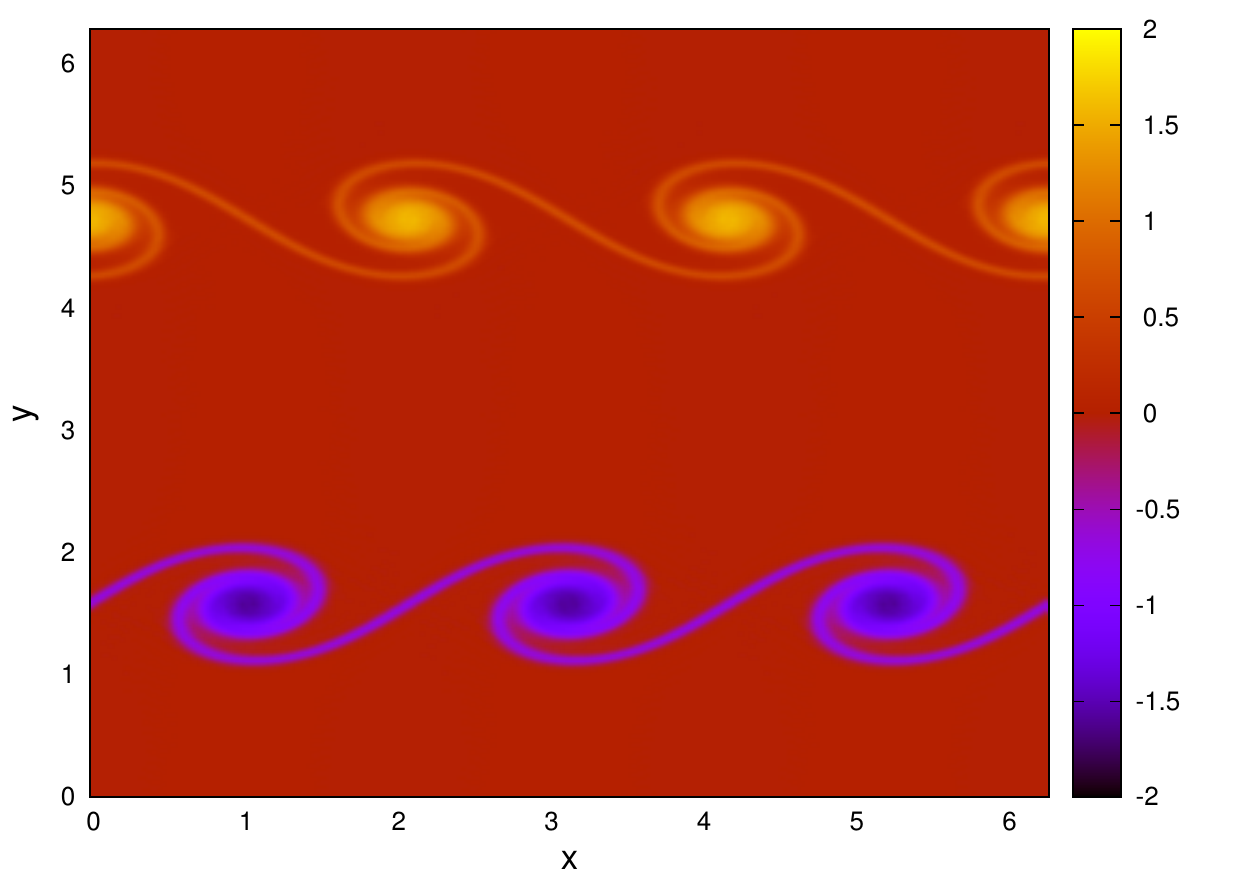}}\\
\subfloat[]{\includegraphics[scale=0.345]{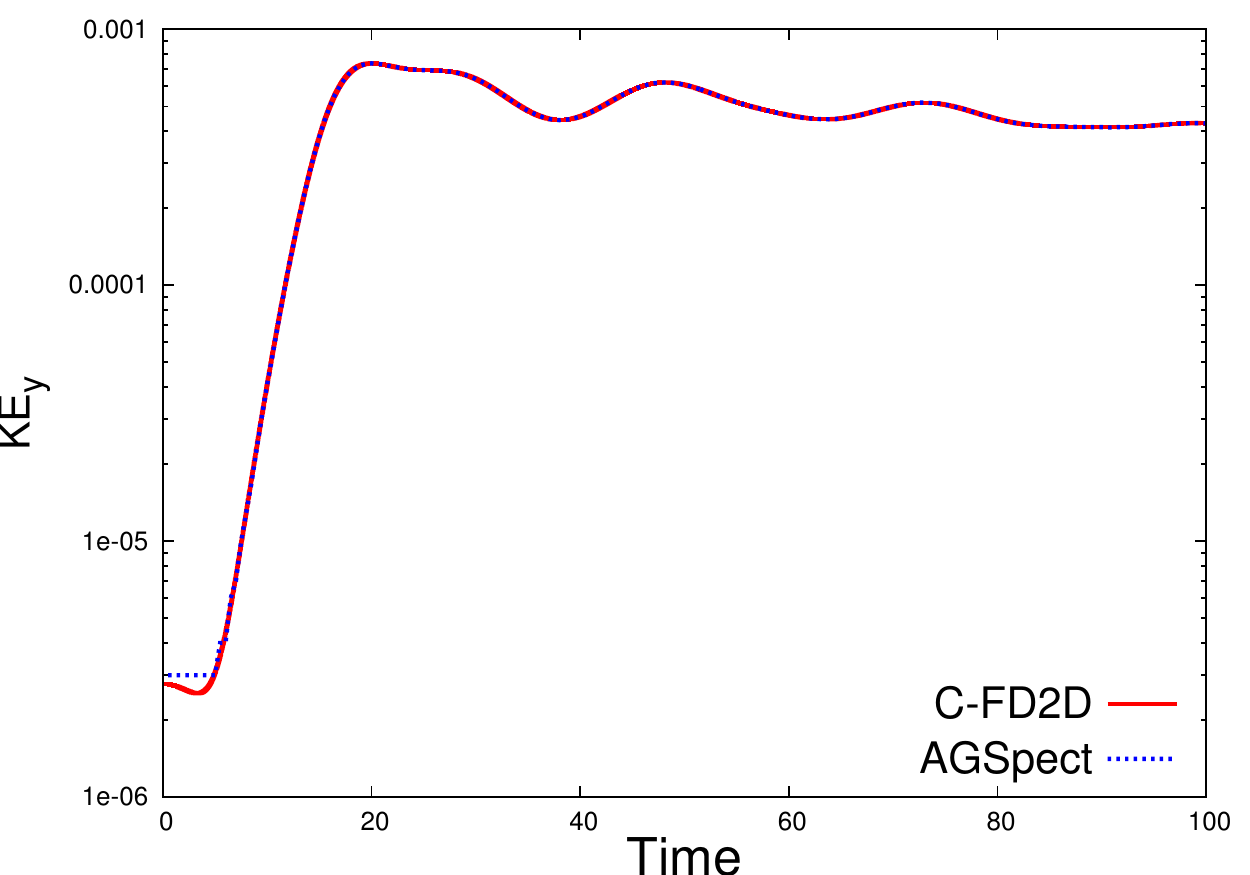}}
\subfloat[]{\includegraphics[scale=0.345]{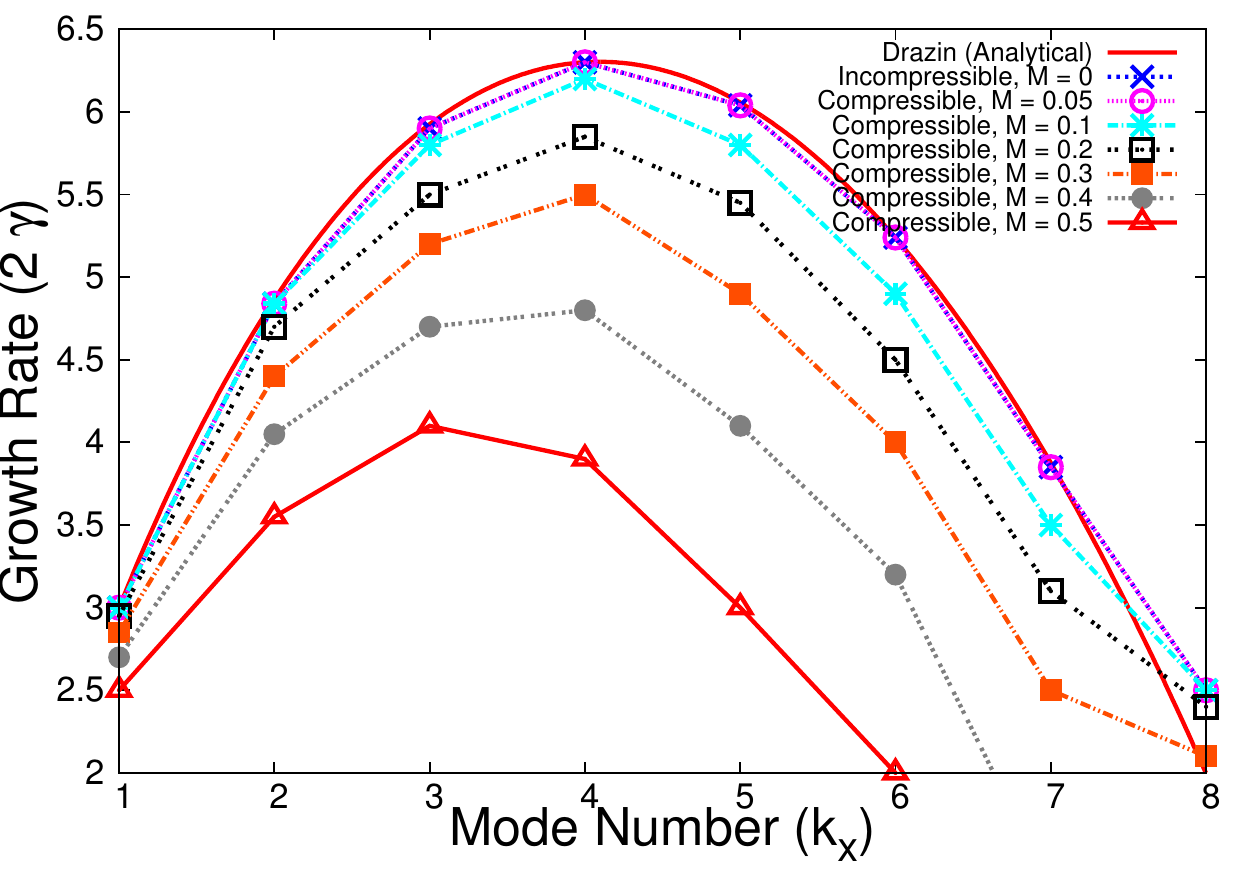}}
\caption{(Color online) (a) Initial velocity profile of K-H instability. (b) Snapshot of vorticity profile of K-H instability for $K_x = 3$ at time $t = 21.9$ with grid resolution $N_x = N_y = 256$, $L_x = L_y = 2 \pi$, $\nu = 10^{-4}$, $dt = 10^{-3}$, $\omega_0 = 2$, $d = \frac{3 \pi}{128}$ and $M = 0$. (c) Growth rate ($2\gamma$) of K-H instability is plotted with mode number ($k_x$) of excitation. The red solid line is evaluated from the analytical expression obtained by Drazin. The blue line with `X' sign represents the growth rate from earlier code in vorticity formalism and the magenta line with symbol `O' represents the same from velocity formalism code with Mach number (M) = 0.05. (d) Comparison of growth rate of K-H instability for $k_x = 3$ with AGSpect \cite{akanksha:2017} for M = 0.05}
\label{drazin}
\end{center}
\end{figure}

In two dimensions, the Navier-Stokes equation in vorticity formalism become,
\begin{eqnarray*}
&& \frac{\partial \vec{\omega}}{\partial t} + \vec{u} \cdot \vec{\nabla} \vec{\omega} = \nu \nabla^2 \vec{\omega} \\
&& \vec{\omega} = \vec{\nabla} \times \vec{u}
\end{eqnarray*}
The Navier-Stokes equation in velocity formalism with compressibility effects become,
\begin{eqnarray*}
&&\frac{\partial \rho}{\partial t} + \vec{\nabla} \cdot (\rho \vec{u}) = 0\\
&&\frac{\partial \vec{u}}{\partial t} + ( \vec{u} \cdot \vec{\nabla} ) \vec{u} = \frac{\mu}{\rho} {\nabla}^2 \vec{u} - \frac{C_s^2}{\rho} \vec{\nabla} \rho
\end{eqnarray*}
In order to simulate the two dimensional hydrodynamic fluid, we time evolve the equations in velocity formalism and reproduce the growth rate of Kelvin-Helmholtz instability for a broken jet in the incompressible limit analytically calculated earlier by Drazin \cite{drazin:1961} and later obtained by Ashwin et al \cite{ashwin:2010} using molecular dynamics simulation. The analytical formula obtained by Drazin \cite{drazin:1961} in our notation becomes,
\begin{eqnarray*}
&& \gamma = \frac{k_x U_0}{3} \left[\sqrt{3} - 2 \frac{k_x}{R_E} -2 \left\{\left(\frac{k_x}{R_E}\right)^2 + 2\sqrt{3} \frac{k_x}{R_E} \right\}^{\frac{1}{2}} \right]\\
&& R_E = \frac{U_0 d}{\nu}
\end{eqnarray*}
where $2 \gamma$ is the growth rate, $k_x$ is the mode number of perturbation, $R_E$ is the Raynold's number, $U_0$ is the maximum fluid velocity, $d$ is the shear width and $\nu$ is the coefficient of viscosity.
We run our simulation with parameters $N_x = N_y = 512; L_x = L_y = 2\pi; \omega_0 = 25; d = \frac{3\pi}{128}; \nu = 1.5 \times 10^{-3}$. The initial velocity profile is obtained from the vorticity profile given by $\omega = \frac{\omega_0}{\cosh^2\left(\frac{y-3L_y/4}{d}\right)} - \frac{\omega_0}{\cosh^2\left(\frac{y-L_y/4}{d}\right)}$. We obtain the identical growth rate from this code with M = 0.05 and another simple code that simulates the Navier-Stokes equation in vorticity formalism with M = 0 [Fig \ref{drazin}c]. We also obtain the same growth rate from another existing, well-benchmarked code AG-Spect \cite{akanksha:2017} at the limit of negligible visco-elasticity ($\tau_m = 10^{-4}$)[Fig \ref{drazin}d]. For this benchmarking the initial condition we have used is $\omega = \frac{\omega_0}{\cosh^2\left(\frac{y+L_y/4}{d}\right)} - \frac{\omega_0}{\cosh^2\left(\frac{y-L_y/4}{d}\right)} + \frac{\omega_0}{\cosh^2\left(\frac{y-3L_y/4}{d}\right)} - \frac{\omega_0}{\cosh^2\left(\frac{y+3L_y/4}{d}\right)}$ with $\omega_0 = 2$ and $k_x = 3$. We also reproduce Kelvin-Helmholtz instability with a velocity gradient and found the same growth rate given by Ray \cite{ray:1982}.\\

\section{Three dimensional numerical tests}

In three dimensions we time evolve the following set of equations 
\begin{eqnarray}
&& \label{density} \frac{\partial \rho}{\partial t} + \vec{\nabla} \cdot \left(\rho \vec{u}\right) = 0
\end{eqnarray}
\begin{eqnarray}
&& \frac{\partial (\rho \vec{u})}{\partial t} + \vec{\nabla} \cdot \left[ \rho \vec{u} \otimes \vec{u} + \left(P + \frac{B^2}{2}\right){\bf{I}} - \vec{B}\otimes\vec{B} \right]\nonumber \\
&& \label{velocity} ~~~~~~~~~ = \mu \nabla^2 \vec{u} + \rho\vec{f}; ~~~ P = C_s^2 \rho\\
%&& \frac{\partial E}{\partial t} + \vec{\nabla} \cdot \left[ \left( E + P + \frac{B^2}{2} \right)\vec{u} - \vec{u}\cdot\left( \vec{B} \vec{B} \right)  - \eta \vec{B} \times \left(\vec{\nabla} \times \vec{B} \right) \right] \\
%&& ~~~~~~~~~~~~~~~~~~~~~~~~~~~~~~~~~~~~~~~~~~~~~~~~~~~~~ = \mu \left(\vec{\nabla} \cdot \vec{u} \right)^2\\
&& \label{Bfield} \frac{\partial \vec{B}}{\partial t} + \vec{\nabla} \cdot \left( \vec{u} \otimes \vec{B} - \vec{B} \otimes \vec{u}\right) = \eta \nabla^2 \vec{B}
\end{eqnarray}

%================================================================

\subsection{for hydrodynamic flow}
In order to simulate hydrodynamic flows we turn off Eq. \ref{Bfield} in the above set of equations. We start with an initial profile $E(k) = A k^4 \exp(-2k^2/k_0^2)$. The RMS (root-mean-square) of the velocity divergence is defined as $\theta^\prime = \langle(\partial u_i/\partial x_i)^2\rangle^{1/2}$ and the Skewness is defined as $S_3 \equiv \frac{\textless \left(\partial u_1/\partial x_1\right)^3\textgreater}{\textless \left(\partial u_1/\partial x_1\right)^2\textgreater^{3/2}}$. We reproduce the time evolution of $\theta^\prime$ and $S_3$ as previously obtained by Samtaney et al .
\begin{figure}[h!]
\begin{center}
\subfloat[]{\includegraphics[scale=0.345]{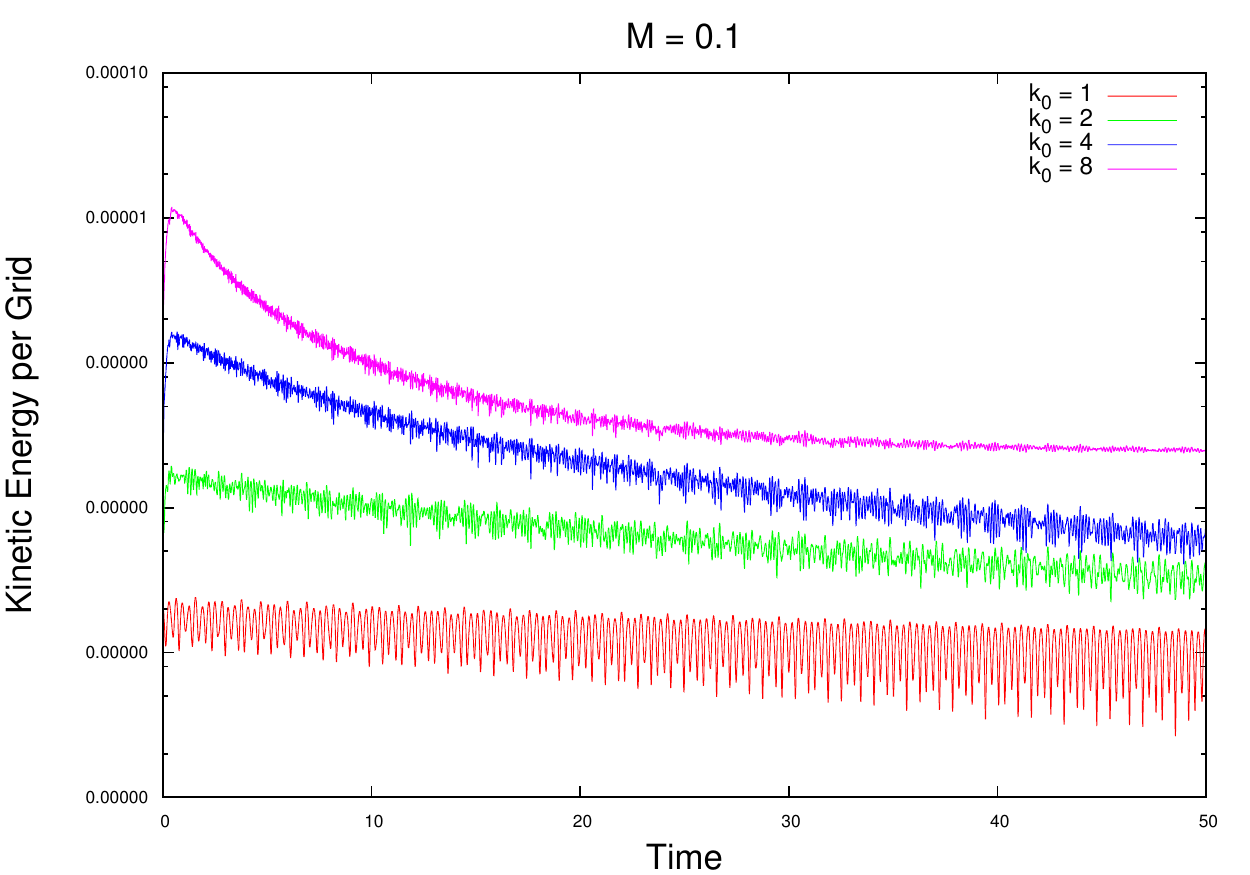}}
\subfloat[]{\includegraphics[scale=0.345]{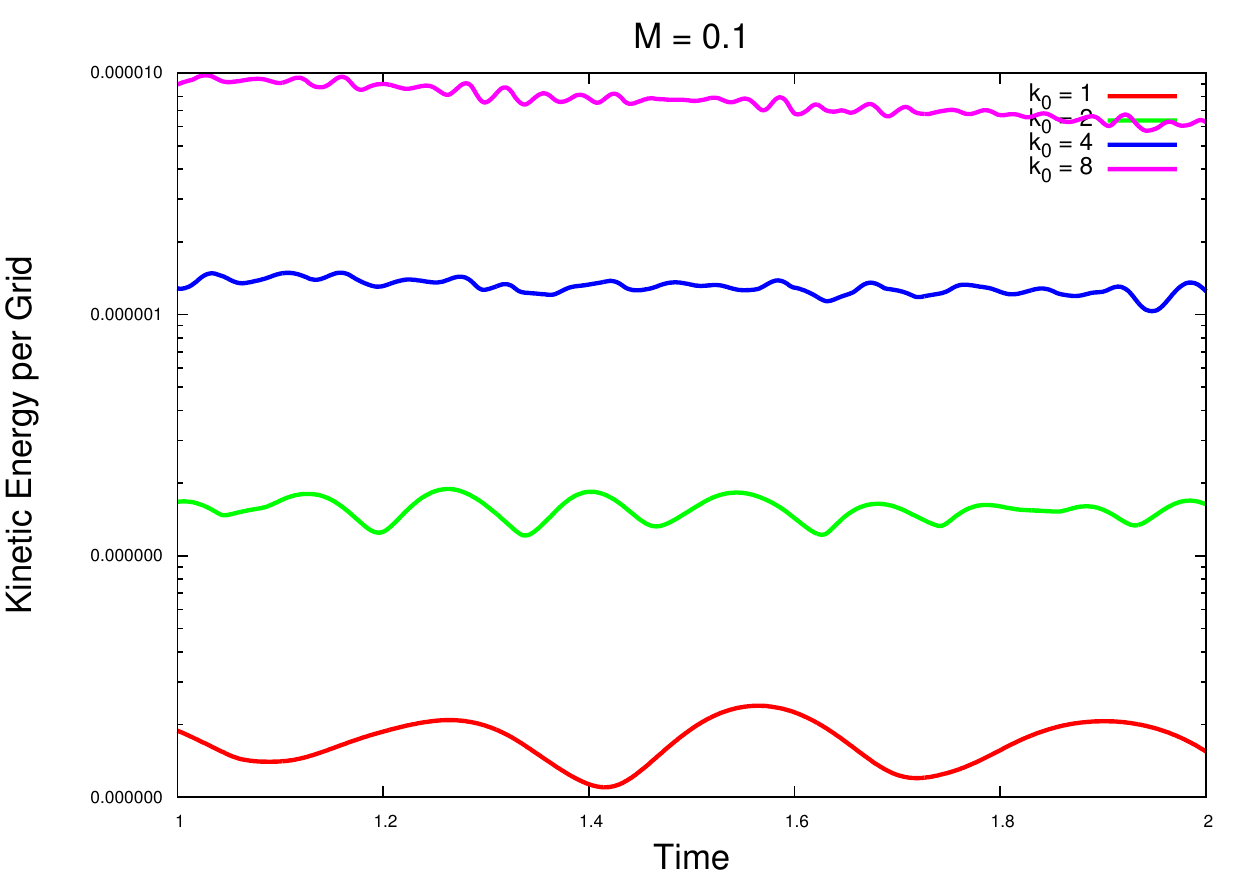}}\\
\subfloat[]{\includegraphics[scale=0.345]{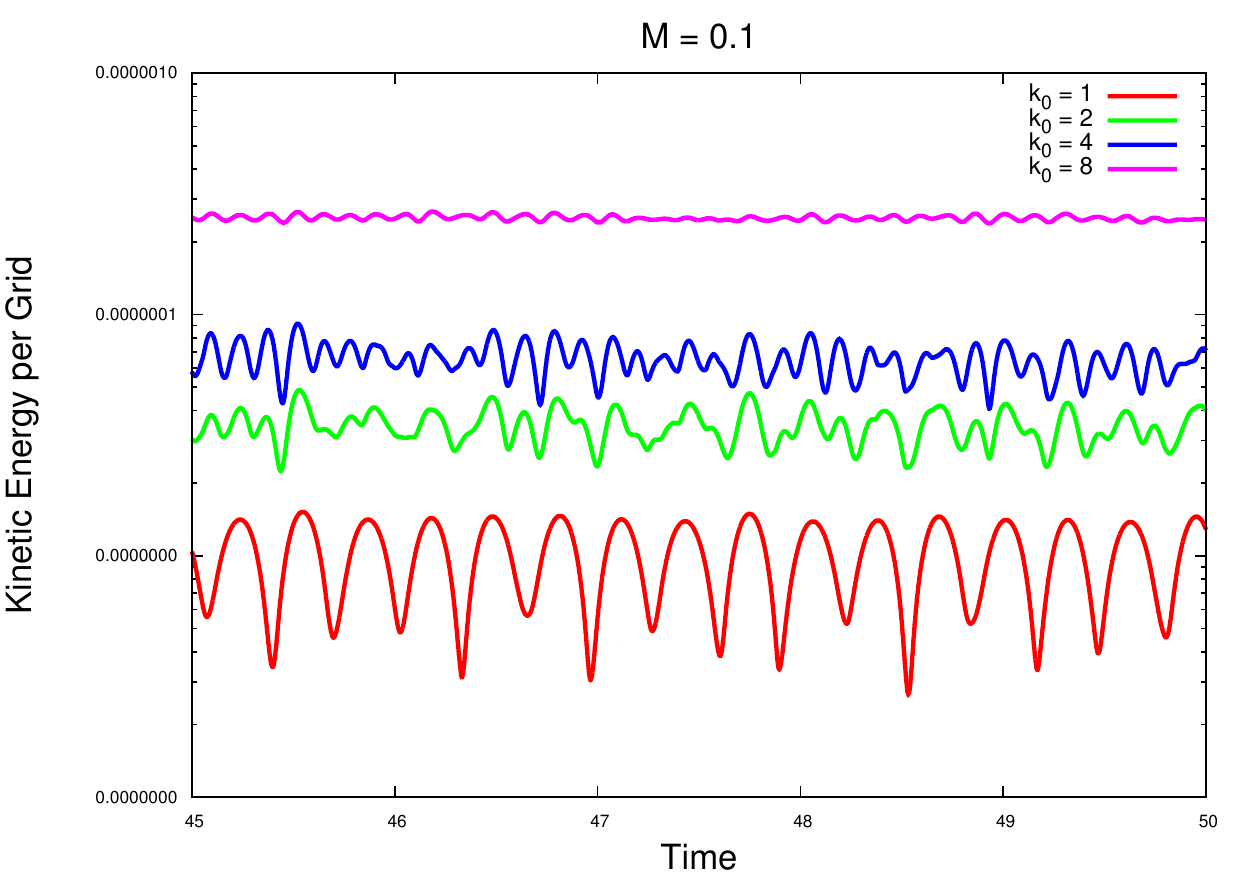}}
\subfloat[]{\includegraphics[scale=0.345]{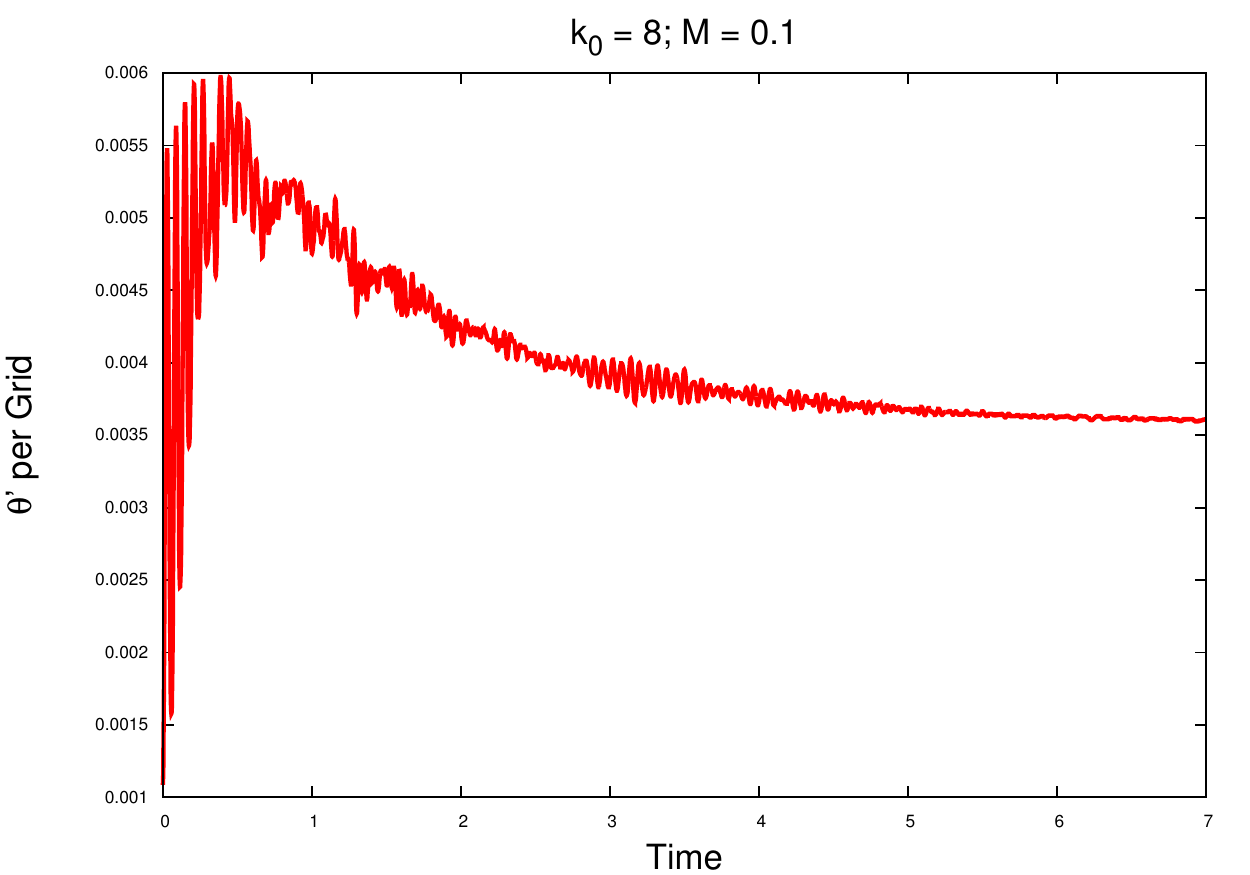}}\\
\caption{(Color online) (a) Time evolution (upto time = 50) of Kinetic energy per grid for different $k_0 (= 1,2,4,8)$ with fixed Mach number $M = 0.1$.  (b) A zoomed view of early time evolution (upto time = 2) of Fig \ref{ravi_samtaney}(a). (c) A zoomed view of late time evolution (upto time = 50) of Fig \ref{ravi_samtaney}(a). (d) Benchmarking of Skewness for $k_0 = 8$ and $M = 0.1$.}
\label{ravi_samtaney}
\end{center}
\end{figure}
\begin{figure}[h!]
\begin{center}
\subfloat[]{\includegraphics[scale=0.345]{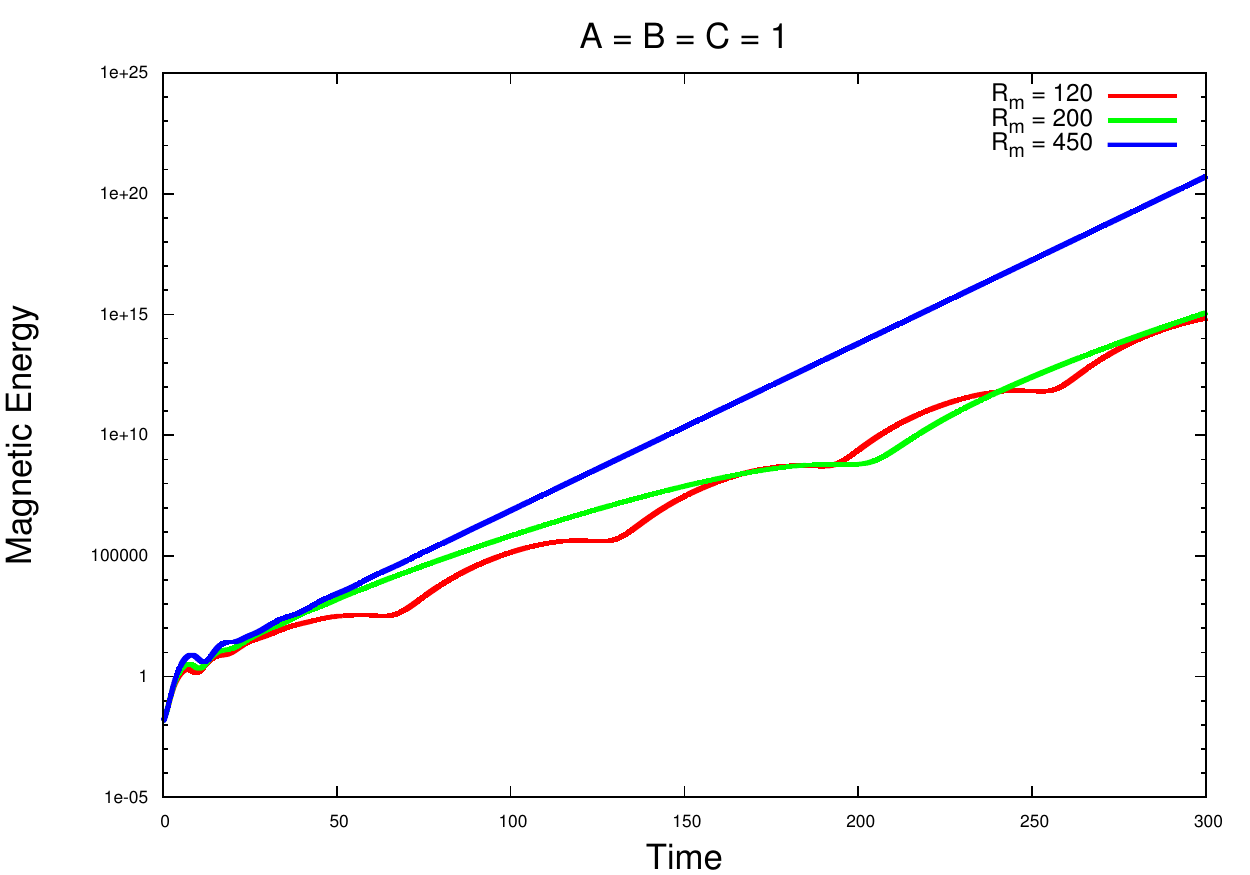}}
\subfloat[]{\includegraphics[scale=0.295]{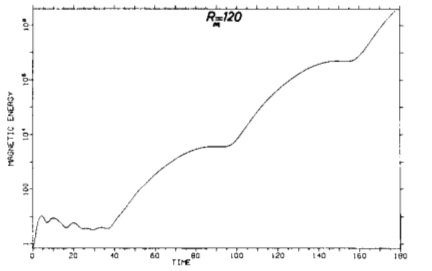}}\\
\subfloat[]{\includegraphics[scale=0.295]{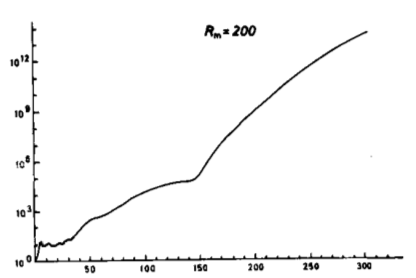}}
\subfloat[]{\includegraphics[scale=0.295]{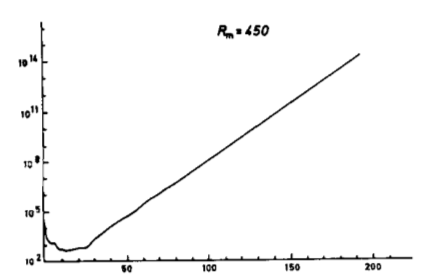}}\\
\subfloat[]{\includegraphics[scale=0.345]{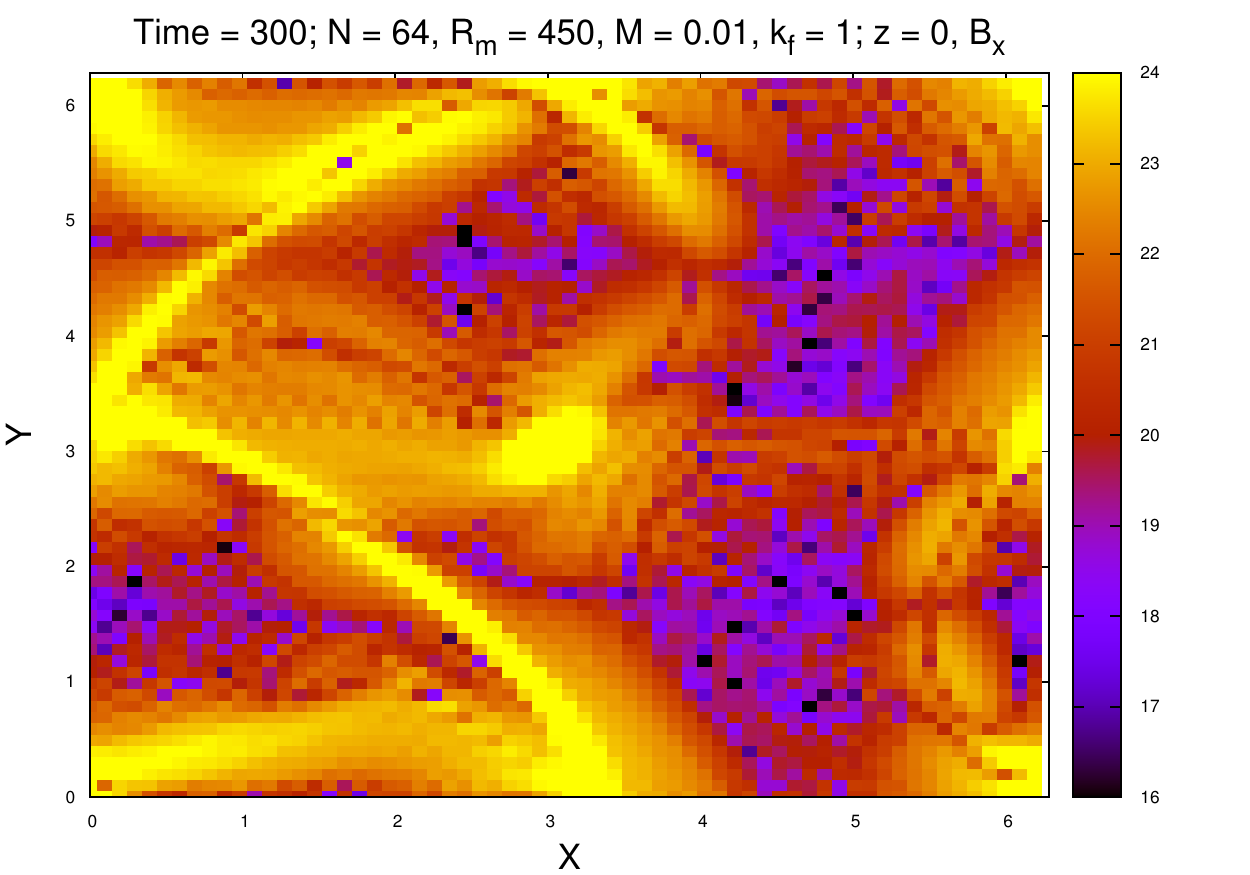}}
\subfloat[]{\includegraphics[scale=0.345]{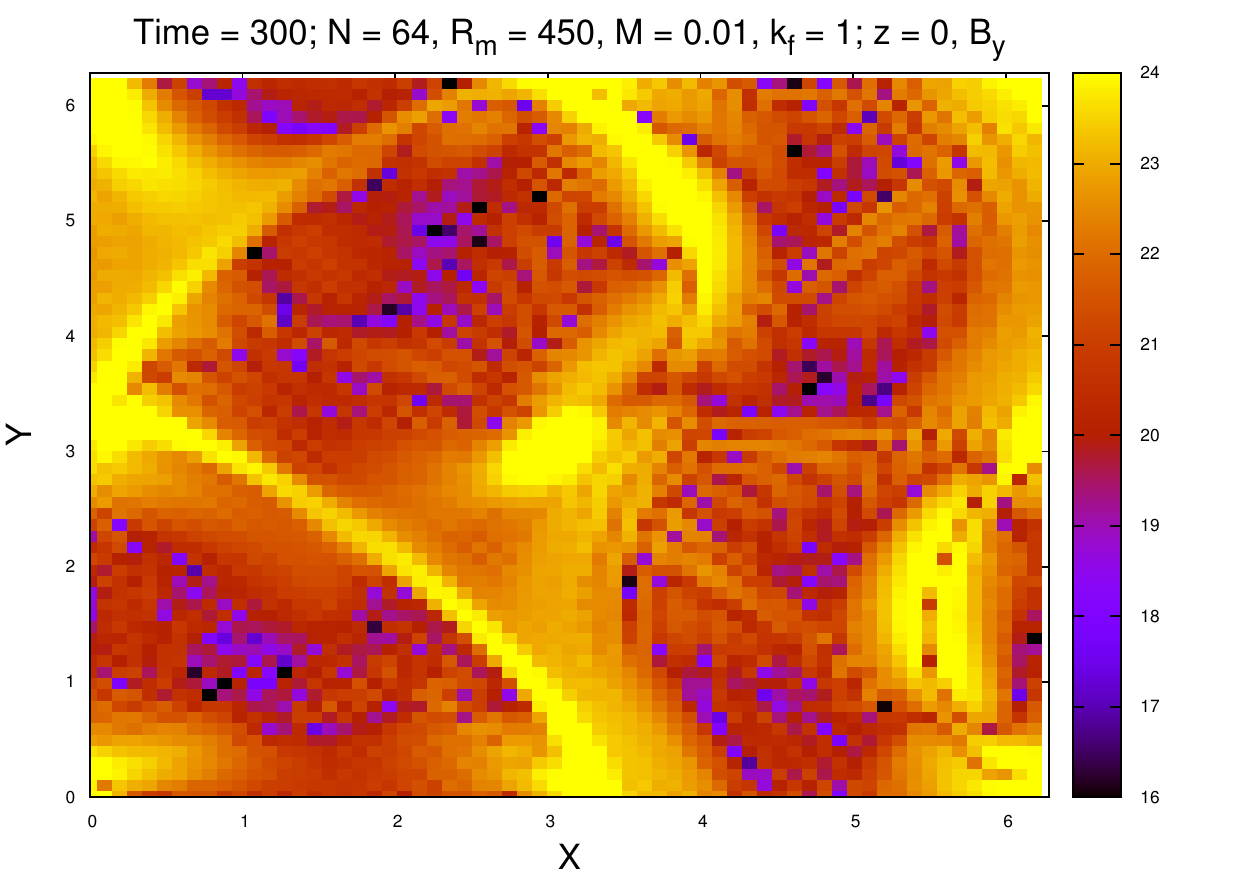}}\\
\subfloat[]{\includegraphics[scale=0.345]{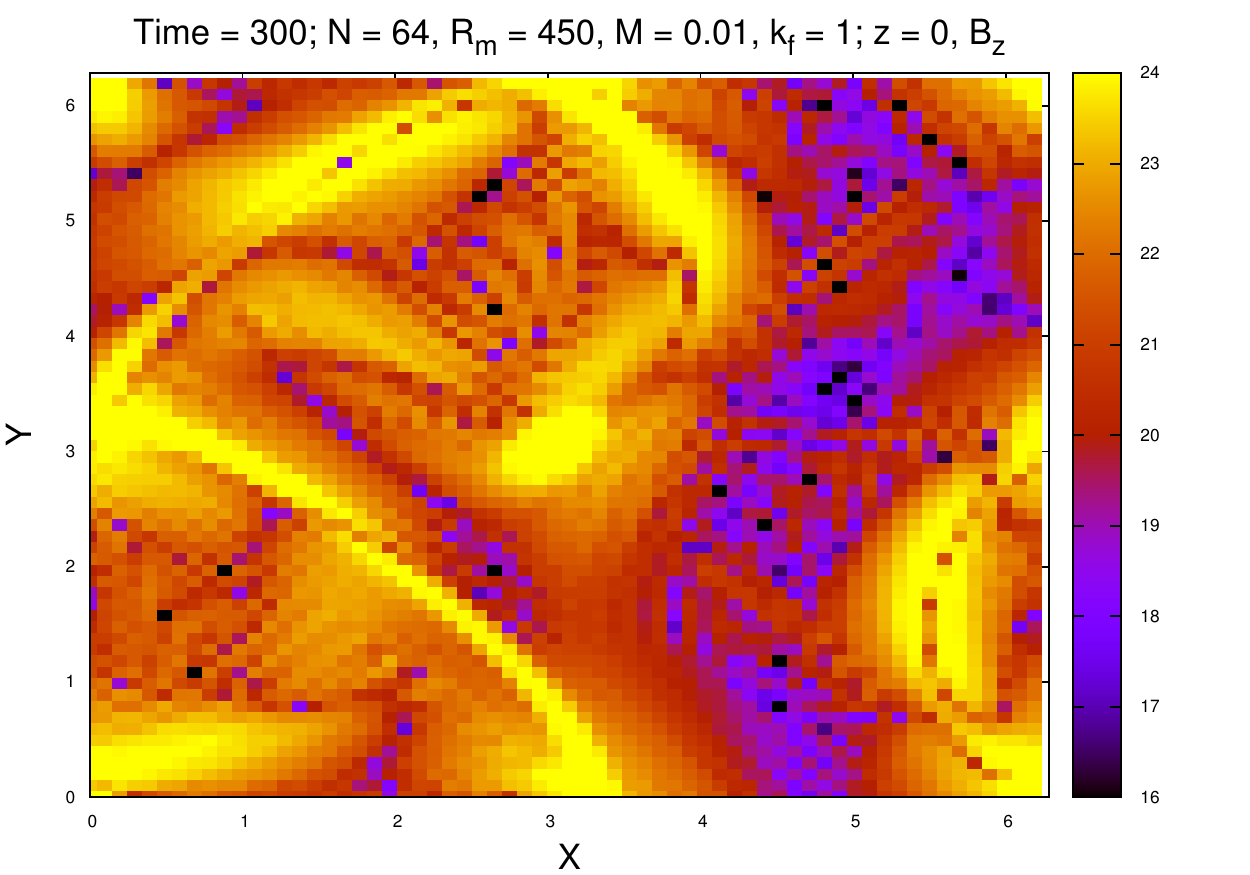}}
\subfloat[]{\includegraphics[scale=0.345]{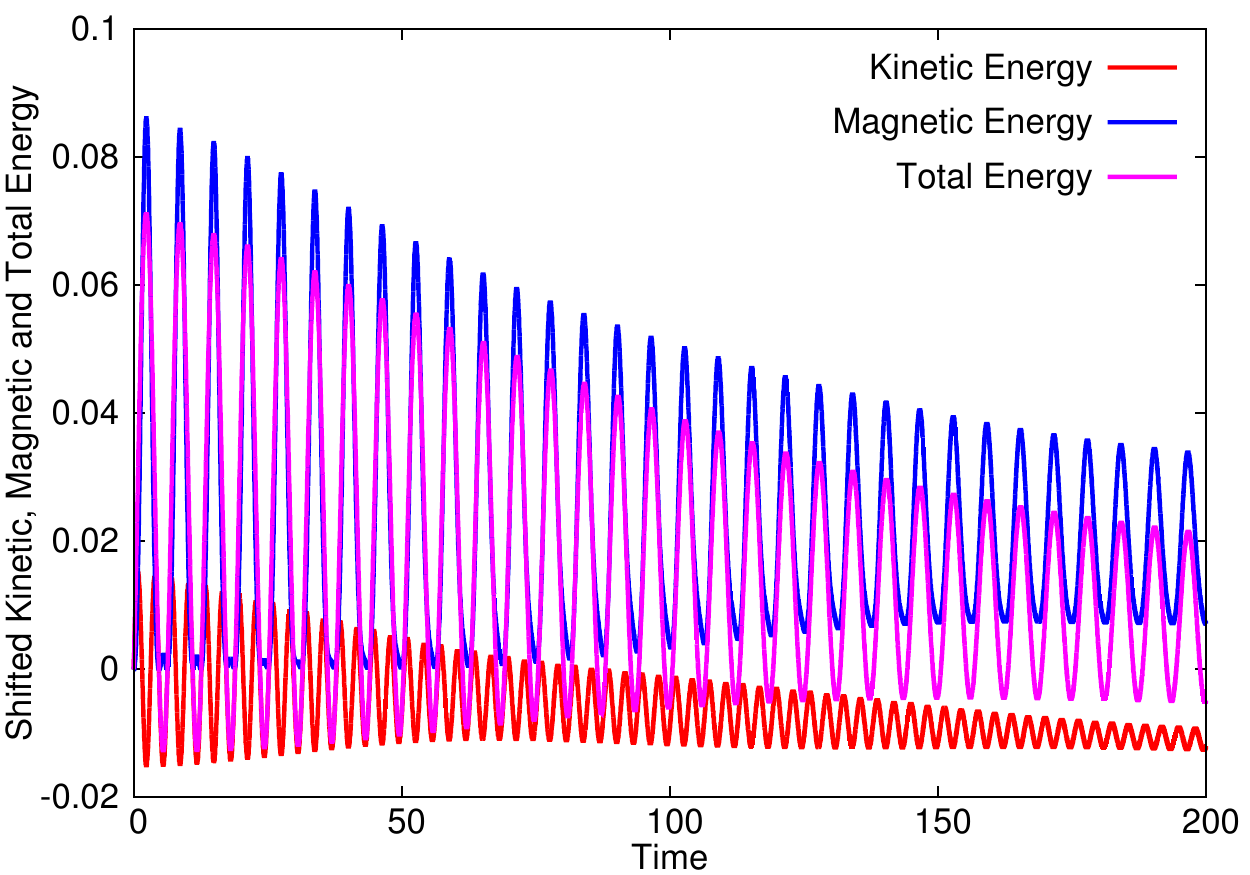}}\\
\caption{(Color online) (a) Comparison of Growth rate of magnetic field for $Rm = 120, 200, 450$ from MHD3D. (b) Growth rate of magnetic field for $Rm = 120$ earlier obtained by Galloway and Frisch. (c) Growth rate of magnetic field for $Rm = 200$ earlier obtained by Galloway and Frisch. (d) Growth rate of magnetic field for $Rm = 450$ earlier obtained by Galloway and Frisch. (e) x-component of magnetic field at mid-plane for grid resolution $64^3$ with $Rm = 450$, $K_f = 1$, $M = 0.01$ at time $= 300$. (f) y-component of magnetic field at mid-plane for grid resolution $64^3$ with $Rm = 450$, $K_f = 1$, $M = 0.01$ at time $= 300$. (g) z-component of magnetic field at mid-plane for grid resolution $64^3$ with $Rm = 450$, $K_f = 1$, $M = 0.01$ at time $= 300$. (h) Time evolution of kinetic, magnetic and total energy per grid for ABC forcing with $k_f = 1$ at $M = 0.1$ and $M_A = 1$. In the presence of external forcing the whole system acts as a forced relaxed system.}
\label{Frisch}
\end{center}
\end{figure}
%================================================================

\subsection{for magnetohydrodynamic flow}
First we turn off Eq. \ref{density} and \ref{velocity} and start with the initial condition $\rho = \rho_0$ as a uniform density fluid, the initial velocity profile as $u_x = U_0 [A \sin(k_f z) + C \cos (k_f y)]$, $u_y = U_0 [B \sin(k_f x) + A \cos (k_f z)]$, $u_z = U_0 [C \sin(k_f y) + B \cos (k_f x)]$ and the initial magnetic field as $B_x = B_y = B_z = B_0$. We choose, $\rho_0 = 1$, $U_0 = 1$, $k_f = 1$ and $A = B = C = 1$ as previously taken by Galloway et al. We reproduce the growth rate of the magnetic energy for the specified ABC flow for Reynold's numbers [Fig. \ref{Frisch}] $Re = 120, 200$ and $450$. The exponential growth of magnetic energy of several orders of magnitude is known as ``dynamo effect".\\

\begin{figure}[h!]
\begin{center}
\subfloat[]{\includegraphics[scale=0.345]{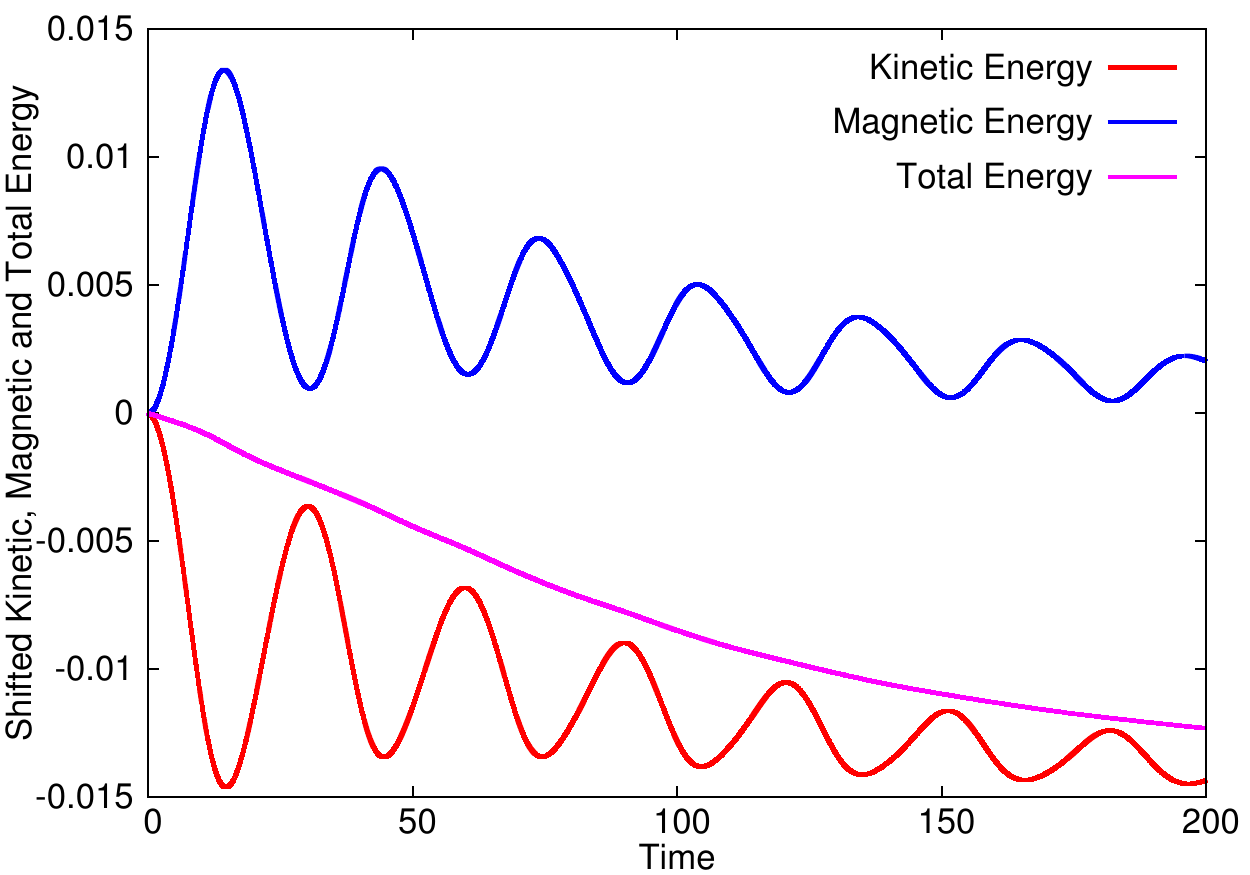}}
\subfloat[]{\includegraphics[scale=0.345]{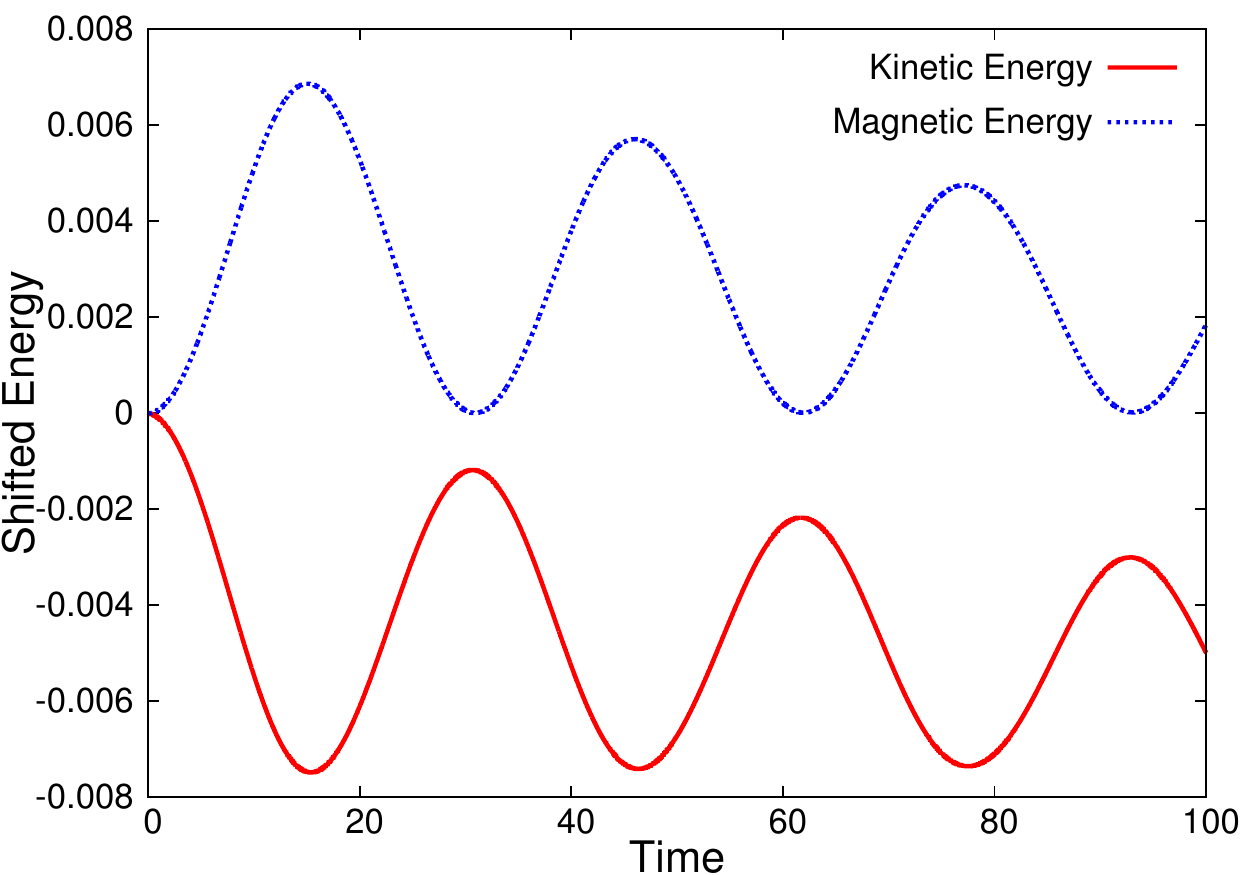}}\\
\subfloat[]{\includegraphics[scale=0.345]{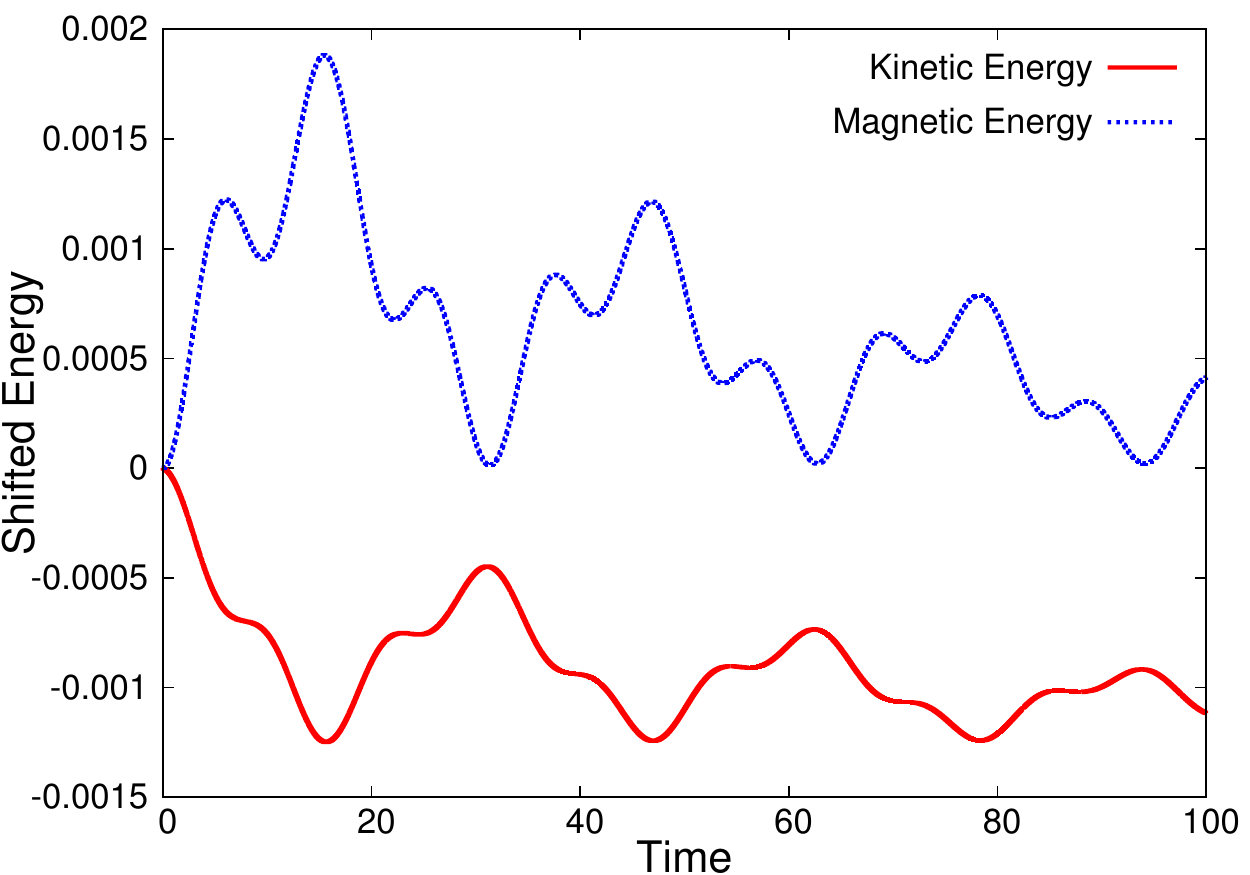}}
\subfloat[]{\includegraphics[scale=0.345]{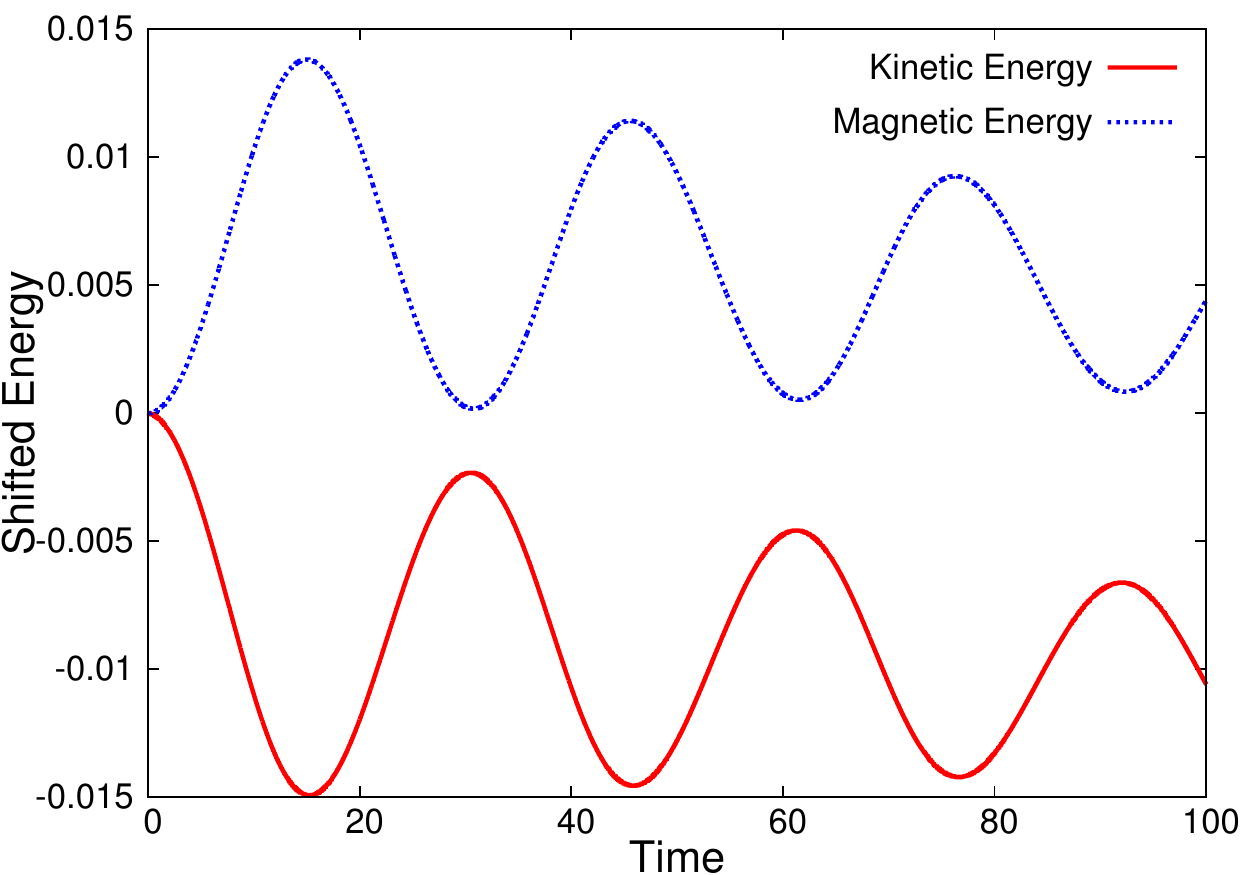}}\\
\caption{(Color online) (a) Conservation of total energy and time evolution of kinetic and magnetic energy for Arnold-Beltrami-Childress flow with $M_A = 1$ and $M = 0.1$ and $Re = Rm = 450$ with $k_0 = 1$.(b) Time evolution of kinetic and magnetic energy for Taylor-Green flow with $M_A = 1$ and $M = 0.1$ and $Re = Rm = 450$ with $k_0 = 1$. (c) Time evolution of kinetic and magnetic energy for Roberts flow with $M_A = 1$ and $M = 0.1$ and $Re = Rm = 450$ with $k_0 = 1$. (d) Time evolution of kinetic and magnetic energy for Cat's Eye flow with $M_A = 1$ and $M = 0.1$ and $Re = Rm = 450$ with $k_0 = 1$.}
\label{energy_exchange}
\end{center}
\end{figure}

Next, we time evolve all the terms viz. Eq. \ref{density}, \ref{velocity} and \ref{Bfield}, in addition with an external forcing defined as [Fig. \ref{Frisch}(h)]: $$\vec{f}_{ABC} = \begin{pmatrix}
A \sin(k_f z) + C \cos (k_f y) \\
B \sin(k_f x) + A \cos (k_f z) \\
C \sin(k_f y) + B \cos (k_f x)
\end{pmatrix}$$ 
In the absence of forcing, the initial conditions we start with are the following:
\begin{eqnarray*}
&& u_x = A \sin(k_0 z) + C \cos (k_0 y),\\
&& u_y = B \sin(k_0 x) + A \cos (k_0 z),\\
&& u_z = C \sin(k_f y) + B \cos (k_f x) \Rightarrow \text{Fig.}\ref{energy_exchange}(a)\\
&& u_x = U_0 [A \sin(k_0 z)], u_y = U_0 [B \sin(k_0 x)],\\
&& u_z = U_0 [C \sin(k_0 y)] \Rightarrow \text{Fig.}\ref{energy_exchange}(b)\\
&& u_x = A ~ U_0 \left[ \cos(kx) \sin(ky) \cos(kz) \right],\\
&& u_y = - A ~ U_0 \left[ \sin(kx) \cos(ky) \cos(kz) \right],\\
&& u_z = 0 \Rightarrow \text{Fig.}\ref{energy_exchange}(c)\\
&& u_x = U_0 [B \sin(k_0 y)], u_y = U_0 [A \sin(k_0 x)], \\
&& u_z = U_0 [A \cos(k_0 x) - B \cos(k_0 y)] \Rightarrow \text{Fig.}\ref{energy_exchange}(d)
\end{eqnarray*}
We choose $M_A = 1$ such that, the initial kinetic and magnetic energies are equal. The kinetic and magnetic energy oscillates by transferring energies between kinetic and magnetic modes keeping the total energy conserved. In Fig.\ref{energy_exchange} the fall of total energy is solely due to viscous and resistive effects. For further check of the accuracy we increase the Alfven Mach number ($M_A$) such that the frequency of the Alfven waves increase leaving its signature in the oscillations of the kinetic and magnetic energy. We choose $M_A = 0.1$ and still found consistant decay of total energy, though some oscillations of tiny amplitude appears.

We also reproduce the kinetic and magnetic energy spectra for an incompressible three dimensional magnetohydrodynamic ABC flow where the backreaction of magnetic field on the velocity fields are not considered. The results were obtained earlier by Sadek {\it et al} \cite{alexakis:2016}.

In order to develop some diagnostics of the 3D MHD code, we calculate the density and velocity structure functions for a initially random velocity and magnetic field. We get back similar structure functions as earlier found by Yang {\it et al} \cite{yang:2017}.

We have developed modules for wavenumber based forcing and gaussian white noise implementation algorithms. We have also incorporated a dynamical wave-number dependent turbulent viscosity model in our code. The coupling in the above set of equations can be altered easily to suit specific purposes.

\section{Tracer Particles}

Tracer particles act as a very important diagnostics for any fluid code. In MHD3D code also, tracer particles are added to discover the nature of turbulence that appears in a MHD plasma. The code evaluates the density, velocity and magnetic fields at the grid points using the MHD description of a plasma. The tracer particles are randomly sprinkled in the simulation domain at the beginning of the simulation. They can be placed anywhere in between the grids or at the grid location. These particles are considered to be passive elements of the code. Thus, they do not contribute to the time evolution of any of the fields that are calculated at the grid location from the MHD equations mentioned in the previous section. But the fields do affect the tracers and determine their motion within the simulation box. The initial velocity of the tracers are evaluated from the velocity at the grid. Next, the time evolution of the tracers are fully governed by the magnetic fields at the grids. The grid magnetic field is interpolated at the position of the tracers in between the grids and the Lorentz force acting on them is evaluated. This force pushes the particles to their new positions. The interpolation of the grid magnetic field is done again at the new positions of the tracers and the new force is evaluated. This procedure is repeated to follow the trajectory of the tracer particles in the plasma. Thus it gives us the information on the motion of the fluid elements in a turbulent plasma and one can follow the deterministic path of a tracer which may reveal more information about the underlying symmetries of the plasma flow. Following we describe the interpolation scheme and the particle pusher algorithm in detail.

\subsection{Interpolation Scheme}
\begin{figure}[h!]
\begin{center}
\includegraphics[width=10cm, height=5cm]{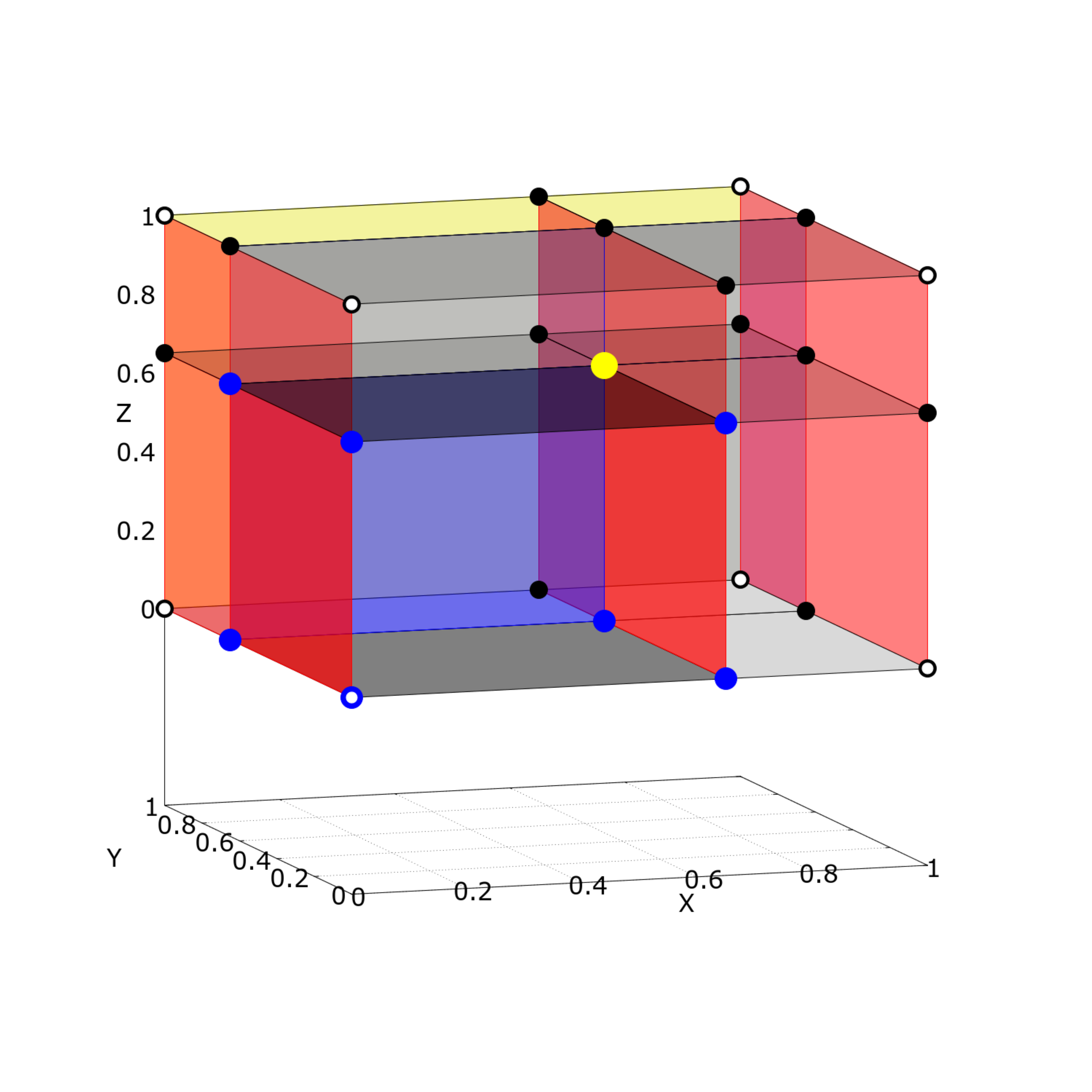}
\caption{(Color online) CIC scheme for volume weighting for the evaluation of magnetic fields on the location of the tracer particles. The tracer particle (yellow filled circle) is located  within the unit shell. It divides the unit shell into 8 subshells, one of them are shown in dark colors with blue circles at the edges. The magnetic field at the tracer particle location ($\vec{B}_p$) is interpolated from the grid magnetic field ($\vec{B}_g$) of the outer corner grids (black hollow circles). The weights of interpolation is given in Eq. \ref{tracer_eq} }
\label{tracer}
\end{center}
\end{figure}

In order to calculate the magnetic force acting on a tracer particle, we use the following Cloud-In-Cell (CIC) scheme. In this scheme, the magnetic field of eight grid points are interpolated at the particle position. In  Fig. \ref{tracer}, we show the volume weighting scheme. The magnetic field is known at all the grid points at the simulaiton domain. The tracer particle divides the unit cube, into eight subshells. One of them (the bottom-left-front) are shown in dark colors (Fig. \ref{tracer}) with blue filled and hollow circles at the corners for better identification. The magnetic field at eight corner-grids (blue and black hollow circles at Fig. \ref{tracer}) are then interpolated at the particle locaton ($\vec{B}_p$) (yellow filled circle at Fig. \ref{tracer}) using the below scheme.\\

If the $m^{th}$ particle is placed within an unit cube whose bottom-left-front corner has a grid index  $(i,j,k)$, the magnetic field acting on the $m^{th}$ particle ($= \vec{B}^m_p$) will be given by, 
\begin{eqnarray}
&& \vec{B}^m_p = \nonumber\\
&& \vec{B}_g^{i,j,k} \left[ (x_g^{i+1}-x_p^m) \cdot (y_g^{j+1}-y_p^m) \cdot (z_g^{k+1}-z_p^m) \right] dt + \nonumber\\
&& \vec{B}_g^{i+1,j,k} \left[ (x_p^m-x_g^{i}) \cdot (y_g^{j+1}-y_p^m) \cdot (z_g^{k+1}-z_p^m) \right] dt + \nonumber\\
&& \vec{B}_g^{i,j+1,k} \left[ (x_g^{i+1}-x_p^m) \cdot (y_p^m-y_g^{j}) \cdot (z_g^{k+1}-z_p^m) \right] dt + \nonumber\\
&& \vec{B}_g^{i+1,j+1,k} \left[ (x_p^m-x_g^{i}) \cdot (y_p^m-y_g^{j}) \cdot (z_g^{k+1}-z_p^m) \right] dt + \nonumber\\
&& \vec{B}_g^{i,j,k+1} \left[ (x_g^{i+1}-x_p^m) \cdot (y_g^{j+1}-y_p^m) \cdot (z_p^m-z_g^{k}) \right] dt + \nonumber \\
&& \vec{B}_g^{i+1,j,k+1} \left[ (x_p^m-x_g^{i}) \cdot (y_g^{j+1}-y_p^m) \cdot (z_p^m-z_g^{k}) \right] dt + \nonumber\\
&& \vec{B}_g^{i,j+1,k+1} \left[ (x_g^{i+1}-x_p^m) \cdot (y_p^m-y_g^{j}) \cdot (z_p^m-z_g^{k}) \right] dt + \nonumber\\
\label{tracer_eq} && \vec{B}_g^{i+1,j+1,k+1} \left[ (x_p^m-x_g^{i}) \cdot (y_p^m-y_g^{j}) \cdot (z_p^m-z_g^{k}) \right] dt
\end{eqnarray}
where, the subscript $p$ denotes the particle and subscript $g$ denotes the grid. $x_g^i, y_g^j, z_g^k$ denote the coordinate of the $(i,j,k)^{th}$ grid. Periodic boundary condition (PBC) has been used such that the tracer particle present at one boundary can feel the magnetic field from the opposite boundary.

%=========================================================================================

\subsection{Particle Pusher Scheme}
We use Boris Algorithm for pushing the particles from its old to new position due to the effect of Lorentz force acting on the particles. This Lorentz force is evaluated by interpolating the grid magnetic fields to the tracer particle location using the CIC scheme mentioned above. If $\vec{x}^m_t$ and $\vec{v}^m_t$ are the position and velocity of $m^{th}$ tracer particle respectively at time $t$, we use the following steps to calculate $\vec{x}^m_{t+dt}$ and $\vec{v}^m_{t+dt}$ at time $t+dt$.
\begin{eqnarray}
&& \vec{t} = \frac{q \vec{B}^m_p}{M} \frac{dt}{2} \nonumber\\
&& \vec{s} = \frac{2\vec{t}}{1+t^2} \nonumber\\
&& \vec{v}_d = \vec{v}_t + \vec{v}_t \times \vec{t} \nonumber\\
&& \label{boris} \vec{v}_{t+dt} = \vec{v}_t + \vec{v}_d \times \vec{s}\\
&& \label{euler} \vec{x}_{t+dt} = \vec{x}_t + \vec{v}_{t+dt} \cdot dt
\end{eqnarray}
where, $q$ and $M$ are the charge and mass of the tracer particle respectively. In this code we work in the normalisations where we assume $q = M = 1$. The particles are guranteed to stay inside the simulation domain by enforcing periodic boundary condition in the position of the tracers.\\

Thus the following steps are executed to determine the dynamics of the tracers in the code.
\begin{enumerate}
\item Determine the grid index ($i, j, k$) of the bottom-left-front corner grid point of the unit cell in which the $m^{th}$ tracer particle is situated.
\item Use Eq. \ref{tracer_eq} to evaluate the magnetic field at the particle position.
\item Calculate the Lorentz force on the $m^{th}$ tracer particle using Eq. \ref{boris}.
\item Determine the new position of the tracer particle using Eq. \ref{euler}
\end{enumerate}

\section{GPU Acceleration}
\label{sec:gpu-acc}

Over the last decade, there has been an increasing use of graphics processing units (GPUs) as general-purpose highly parallel computing units. 
The GPU is a massively parallel computing device consisting of a large number of streaming multiprocessors (SM), each of which is a set of computing cores. 
Each SM has a fixed number of registers and a fast on-chip memory that is shared among its cores. The different SMs share a slower off-chip memory called the device memory, which is 
much larger in capacity. For example, the Tesla P100 GPU has 56 SMs each with 64 cores, resulting in a total of 3584 cores. Each SM of the Tesla P100 has a 256 KB register file size and 
64 KB of fast on-chip memory. All SMs share a 4 MB L2 cache and 16 GB of device memory. 

GPGPU computing refers to general purpose computing on graphics processing
units (GPUs). In this context, a GPU is typically called a device, while the CPU is called a host. A kernel refers to a program that is executed on the GPU device using a large number
of threads. 

OpenACC~\cite{OpenACC} is an approach for quick and easy porting of existing applications from CPU to GPU. The next section introduces the OpenACC programming paradigm 
and describes how the MHD3D code was accelerated using the OpenACC pragmas.

\subsection{OpenACC Acceleration of MHD3D}
\label{sec:openacc}

OpenACC is a compiler directive-based programming model, designed to facilitate quick porting of existing applications to accelerators like GPUs, without significant programming effort.
OpenACC directives, which are pragmas in C/C++ and specialized comments in Fortran, provide the compiler with additional information, enabling them to optimize the code within the 
directives for a specific accelerator. With openACC, the same code can be compiled for different accelerators, thereby providing performance portability. 

OpenACC provides two types of directives broadly - data directives and compute directives. The data directives provide hints to the compiler on movement of data between the host memory 
and accelerator memory. The compute directives define regions of the host program that need to be compiled as accelerator code. More details on the OpenACC directives can be found in the 
OpenACC Programming and Best Practices Guide~\cite{OpenACCprogguide} and the OpenACC specifications document~\cite{OpenACCspecs}.   

The MHD3D code has triply nested loops interspersed with fourier transforms. To port the triply nested loop regions to the GPU, we use the OpenACC parallel and loop constructs coupled with the collapse clause~\cite{OpenACCspecs}. %Figure~\ref{fig:openaccloop} shows how the OpenMP loop regions are mapped to OpenACC parallel regions.

%\begin{figure}
%\centering
%\frame{\includegraphics[scale=0.38]{openaccloop.png}}
%\caption{Mapping of OpenMP loops in MHD3D to OpenACC parallel loops in G-MHD3D.}
%\label{fig:openaccloop}
%\end{figure}

The compiler that supports OpenACC, compiles the loop region into a GPU kernel and distributes the iteration space of the triply nested loop between threads on the GPU. There are clauses
that the programmer can use to tune how the loops are distributed amongst the threads. 

\begin{figure}
\centering
\frame{\includegraphics[scale=0.45]{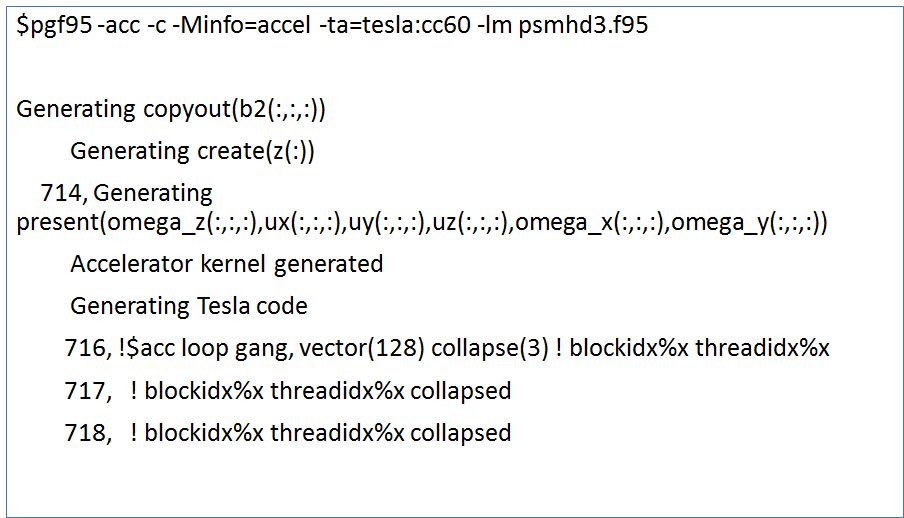}}
\caption{OpenACC Compilation of G-MHD3D.}
\label{fig:openacc-compile}
\end{figure}

Figure~\ref{fig:openacc-compile} is a snapshot of the compile command and the compiler output. The -acc flag tells the compiler to process the openACC directives, -ta=tesla:XX specifies
the architecture for which the accelerator code should be compiled and -Minfo=accel makes the compiler output information corresponding to all the accelerator regions in the code.
The compiler output tells the user which loops are transformed into GPU kernels and how the threads are organized. Information about what data is copied into or out of the accelerator is
also dumped. 

\subsection{cuFFT Library}
\label{sec:cufftlib}

\begin{figure}
\centering
\frame{\includegraphics[scale=0.35]{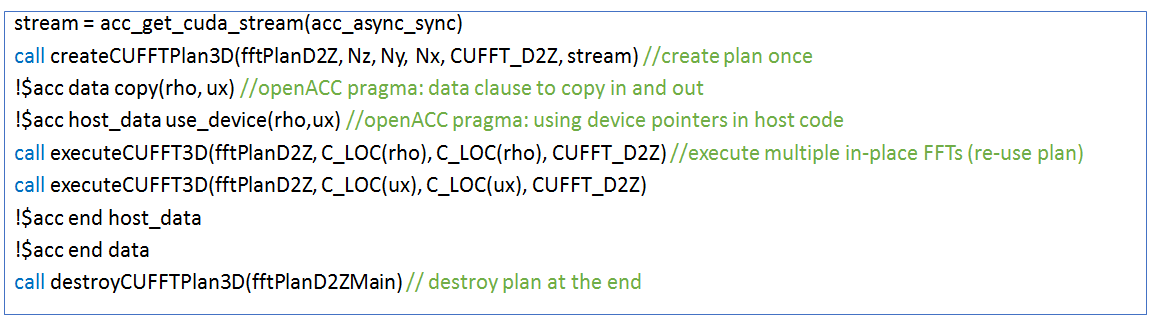}}
\caption{Steps to call cuFFT routines in G-MHD3D.}
\label{fig:cufftex1}
\end{figure}

\begin{figure*}
\centering
\frame{\includegraphics[width=18cm,height=5cm]{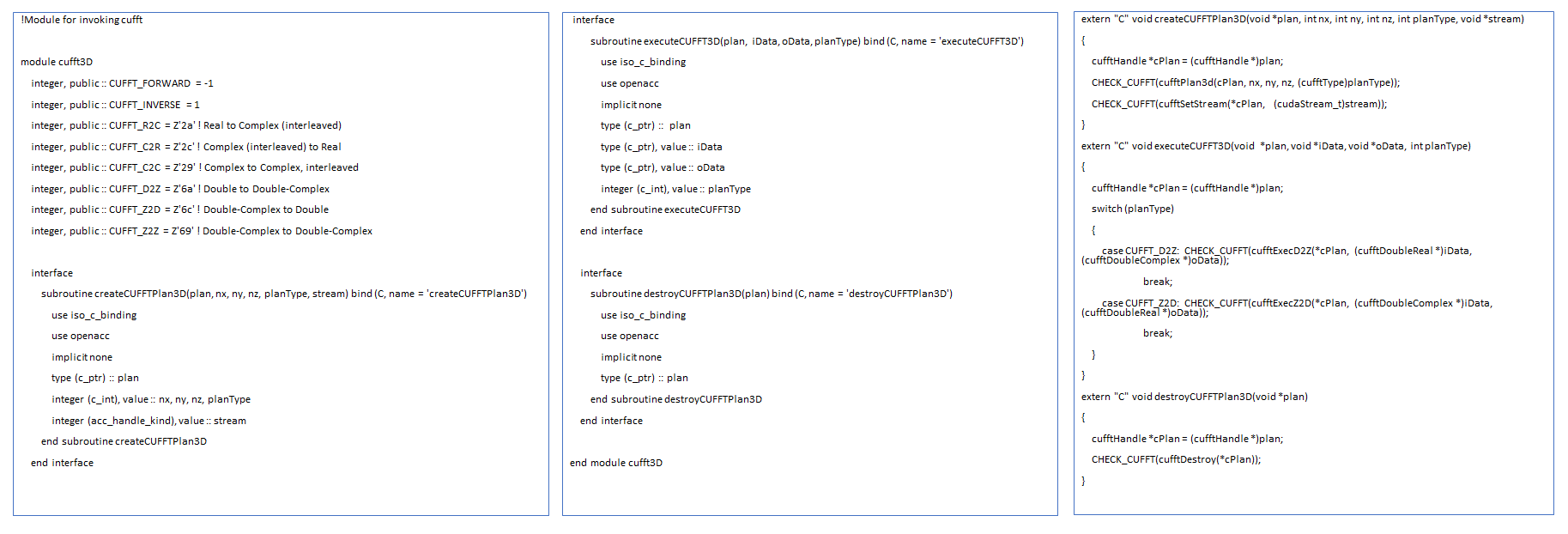}}
\caption{Calling cuFFT routines through C wrapper modules in G-MHD3D.}
\label{fig:cufftex2}
\end{figure*}

The MHD3D code, as mentioned earlier, has loop regions interspersed with fourier transforms. To accelerate the fourier transforms on the GPU, we call routines of the cuFFT library. The 
NVIDIA cuFFT library~\cite{cuFFT} APIs operate on data in the GPU device memory. Hence, before calling the cuFFT routines, we need to move the relevant data to the GPU, if not already 
present, and pass the device pointers to the API calls. The cuFFT plan creation (this is needed for initializing some state information that is necessary before calling the fft routines)
is done once in the beginning and destroyed at the end. Figure~\ref{fig:cufftex1} shows the steps needed to call the cuFFT routines. To call the cuFFT library routines from our OpenACC 
Fortran application, we create C wrapper modules as shown in Figure~\ref{fig:cufftex2}. In-place transforms are used, i.e the output transforms are written to the input arrays. This 
reduces the memory footprint (amount of memory used by the application). 

\subsection{Out-of-core Processing}
\label{sec:outofcore}

While computing the MHD3D equations for large 3D grids of size 512x512x512, we observe that the data operated on within the triply nested loops (aka working set) exceeds the 
device memory capacity of the GPU. Each variable array in this case is 1 GB in size. This results in the need for efficient out-of-core processing. Out-of-core processing refers 
to processing of data that does not completely fit within the memory of a computing system. One approach to out-of-core processing is to split the data into blocks (also called tiles),
 move one tile to the GPU memory and completely process it  before moving it back to the host memory. However, since the solver code is interspersed with fourier transforms that need 
the complete data, it is not possible to do a tile-by-tile processing.  

For seamless out-of-core processing without programmer intervention, in G-MHD3D, we leverage NVIDIA's unified memory~\cite{umemory}.  Unified memory is a component of NVIDIA's GPU programming model that defines a managed memory space in which all processors (CPU and GPU) see a single coherent memory image with a common address space. This means that there is no need to
explicitly move and manage data between the host memory and GPU. For example, when a variable is accessed on the GPU within a kernel and the page that holds the variable is not present 
on the GPU device, (a page is a contiguous block of virtual memory), a page fault occurs. The system transparently handles this and moves the page to the GPU device if necessary. This is 
called on-demand page migration.   

Unified memory also facilitates over-subscription of the GPU memory. This means that you can use variables whose sizes exceed the physical memory capacity of the GPU  and the system 
transparently manages moving parts of data between host and GPU memory as needed. 

To enable unified memory, you need to compile and link your application with the -ta:tesla=managed flag. 

\section{Performance Results}
In this section, we present the performance analysis of the GPU accelerated 3D compressible magnetohydrodynamic code on different GPU architectures. The accelerated code, as specified in
Section~\ref{sec:gpu-acc}, uses OpenACC~\cite{OpenACC} for GPU parallelization. In-place fourier transforms are computed on the GPU using the cuFFT library~\cite{cuFFT}. Computing the 
transforms in-place reduces the memory footprint, resulting in reduced data transfers between the host and GPU and improved performance. For seamless out-of-core processing of large grids 
that do not fit in the GPU memory, we leverage NVIDIA's unified memory~\cite{umemory}. 

Experiments are run on a Intel Xeon E5-2698 v3 16 core, dual socket CPU @2.3 GHz, 256 GB RAM and the NVIDIA
Tesla K80 (Kepler) GPU, NVIDIA Tesla P100 (Pascal) and NVIDIA Tesla V100 (Volta) GPUs. We use cuda toolkit 9.0.176 and pgi 17.9. Table~\ref{tbl:G-MHD3D-perf1} shows the execution time of the solver per timestep for grid sizes of 64X64X64 and 128X128X128 on the CPU and the Kepler, Pascal and Volta GPUs. The CPU version of the code is developed using OpenMP parallelization and the FFTW library. We see that the GPU code is about two orders of magnitude faster than the CPU code.
 
Across GPU architectures, we see that Tesla P100 is 2x to 3x faster than the Tesla K80. This is due to the difference in the memory bandwidth between the two architectures. Tesla P100 has a theoretical peak memory bandwidth of 720 GB/s which is 3x that of Tesla K80 whose peak memory bandwidth is 240 GB/s. As the MHD3D code is memory intensive, we can see 2x-3x performance difference between the two architectures. Performance results over the Pascal and Volta architecture are quite similar (within 30\%) as there is not a great difference in their memory bandwidths (720 GB/s vs 900 GB/s). To verify the correctness of the GPU accelerated solver, we compared the CPU and GPU outputs and computed the mean square error. This was found to be of the order of 5.89E-15.

\begin{table}[h]
\begin{tabular}{|c|c|c|c|c|}
\hline
Grid Size & CPU & Kepler K80 & Tesla P100 & Tesla V100 \\
\hline
64X64X64 & 0.4 & 0.02 & 0.01 & 0.01 \\ 
128X128X128 & 5.89 & 0.12 & 0.04 & 0.03 \\
\hline 
\end{tabular}
\caption{Execution time per timestep in seconds over CPU and Kepler, Pascal and Volta GPUs.}
\label{tbl:G-MHD3D-perf1}
\end{table}

For larger grid sizes which do not fit in GPU memory, we use unified memory. On demand page migration and over-subscription features of unified memory are not supported on the Kepler K80 architecture. Hence, our analysis is restricted to the Pascal and Volta architectures.

\begin{table}[h]
\begin{tabular}{|c|c|c|c|}
\hline
Grid Size & CPU & Tesla P100 & Tesla V100 \\
\hline
256X256X256 & 65.82 & 0.32 & 0.22  \\ 
512X512X512 & 566.12 & 80.97 & 92.81 \\ 
\hline
\end{tabular}
\caption{Execution time per timestep in seconds over CPU, Pascal and Volta GPUs for large grids.}
\label{tbl:G-MHD3D-perf2}
\end{table}

Table~\ref{tbl:G-MHD3D-perf2} shows the execution time per timestep for large grids. For 256X256X256 grid size, the total memory footprint of the solver exceeds the GPU device memory of 16 GB. However, the working set of each kernel is within the GPU memory capacity. When the grid size becomes 512X512X512, the working set of the GPU kernels is in some cases ~40GB and does not fit completely within the GPU memory. As mentioned in Section~\ref{sec:outofcore}, tiling the data typically helps in such a scenario, but since the solver code is interspersed with fourier transforms that need the complete data, it is not possible to do a tile-by-tile processing. 

Therefore, in this case, there is frequent movement of pages back and forth between the CPU and GPU and a large number of page faults on the
GPU. To reduce the page faults on the GPU (GPU page fault handling is expensive), we ensure that the data arrays on the CPU are pre-mapped on the GPU. This avoids a page fault on the first
access. Eventually the page may migrate to the GPU (based on the unified memory system heuristics), but the mappings on the GPU page tables are kept updated.

Even with the above optimizations, the out-of-core solver is an order magnitude slower than the in-core solver, largely due to movement of pages between host and GPU and handling of
the GPU page faults.  Even with this additional overhead, the GPU accelerated code is able to show a speedup of 7x over the OpenMP version.

\section{Future Scope}
The implementation of some diagnostics viz. field line tracer, poincare section, modo-mode energy transfer in a triad \cite{verma:2004} is in progress. The performance of the code can be improved by implementing implicit time solvers. In order to further enhance the speed of the code, the GPGPU parallel cuda version of the code is being developed. At present the code can handle only periodic boundaries. The simulation of plasma flows in bounded domains with upgraded version of the code is under development.


\section*{Acknowledgment}
The development as well as benchmarking of MHD3D has been done at Udbhav and Uday clusters at IPR. For benchmarking and performance results of G-MHD3D, K80, P100 and V100 GPU cards of NVIDIA has been used. RM thanks Samriddhi Sankar Ray, Akanksha Gupta, P N Maya, Mahendra K Verma and Harishankar Ramachandran for several helpful discussions. RM acknowledges the support from ICTS program: ICTS/Prog-dcs/2016/05.



%\bibliographystyle{IEEEtran}

\end{document}